\documentclass{article}

\usepackage{arxiv}

\usepackage[numbers,sort&compress]{natbib}

\usepackage[utf8]{inputenc} % allow utf-8 input
\usepackage[T1]{fontenc}    % use 8-bit T1 fonts
\usepackage{hyperref}       % hyperlinks
\usepackage{url}            % simple URL typesetting
\usepackage{booktabs}       % professional-quality tables
\usepackage{amsfonts}       % blackboard math symbols
\usepackage{nicefrac}       % compact symbols for 1/2, etc.
\usepackage{microtype}      % microtypography
\usepackage{xcolor}         % colors
\usepackage{amsmath}
\usepackage{graphicx}
\usepackage{algorithm}
\usepackage{algpseudocode}
\usepackage{subcaption}
\usepackage{amsfonts}       % blackboard math symbols
\usepackage{newunicodechar}
\usepackage{array}
\title{Multiscale Physics-Informed Neural Network for Complex Fluid Flows with Long-Range Dependencies}

  \author{
 Prashant Kumar \\
  Department of Aerospace Engineering\\
  Indian Institute of Technology Kanpur\\
  Kanpur, India 208016 \\
  \texttt{kumarp22@iitk.ac.in} \\
  \And
 Rajesh Ranjan \\
  Department of Aerospace Engineering\\
  Indian Institute of Technology Kanpur\\
  Kanpur, India 208016 \\
  \texttt{rajeshr@iitk.ac.in} \\
}

\begin{document}

\maketitle

\begin{abstract}
Fluid flows are governed by the nonlinear Navier-Stokes equations, which can manifest multiscale dynamics even from predictable initial conditions. Predicting such phenomena remains a formidable challenge in scientific machine learning, particularly regarding convergence speed, data requirements, and solution accuracy.
In complex fluid flows, these challenges are exacerbated by long-range spatial dependencies arising from distant boundary conditions, which typically necessitate extensive supervision data to achieve acceptable results. We propose the Domain-Decomposed and Shifted Physics-Informed Neural Network (DDS-PINN), a framework designed to resolve such multiscale interactions with minimal supervision. By utilizing localized networks with a unified global loss, DDS-PINN captures global dependencies while maintaining local precision. The robustness of the approach is demonstrated across a suite of benchmarks, including a multiscale linear differential equation, the nonlinear Burgers' equation, and data-free Navier-Stokes simulations of flat-plate boundary layers. Finally, DDS-PINN is applied to the computationally challenging backward-facing step (BFS) problem; for laminar regimes ($Re=100$), the model yields results comparable to computational fluid dynamics (CFD) without the need for any data, accurately predicting boundary layer thickness, separation, and reattachment lengths. For turbulent BFS flow at $Re=10,000$, the framework achieves convergence to $\mathcal{O}(10^{-4})$ using only 500 random supervision points ($<0.3\%$ of the total domain), outperforming established methods like Residual-based Attention-PINN in accuracy. This approach demonstrates strong potential for the super-resolution of complex turbulent flows from sparse experimental measurements.
\end{abstract}


\section{Introduction} \label{sec:intro}
\begin{figure}
    \centering
    \includegraphics[width=0.75\linewidth]{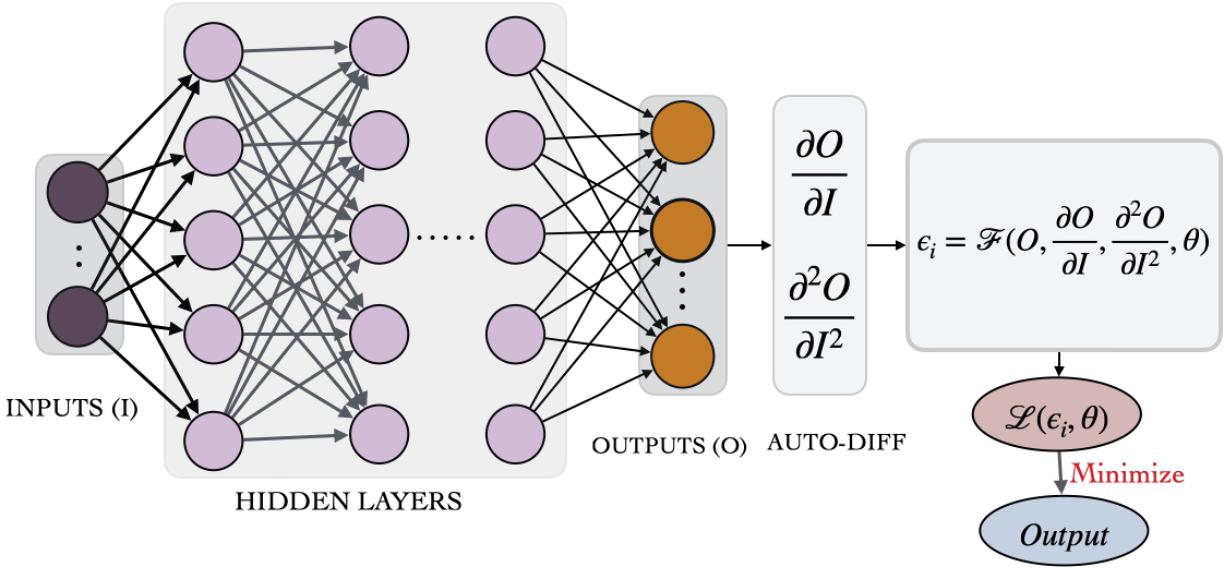}
    \caption{Basic PINN architecture for solving PDEs}
    \label{fig:basic_pinn_arch}
\end{figure}

Physics-Informed Neural Networks (PINNs; ~\cite{raissi2017physics2,raissi2020hidden})represent a transformative paradigm in scientific machine learning, seamlessly integrating data-driven modeling with the rigorous enforcement of physical laws. Figure~\ref{fig:basic_pinn_arch} illustrates a typical feedforward PINN architecture applied to partial differential equations (PDEs), where a sequence of hidden layers transforms input  ($I$) into predicted physical fields ($O$). Through back-propagation, the network parameters ($\theta$) are optimized to minimize a composite loss function $\mathcal{L}(\epsilon_i,\theta)$, which enforces the governing PDEs as a residual constraint $\epsilon_i$ at discrete collocation points. This process ensures that the neural network's approximation remains consistent with the underlying physical laws while simultaneously fitting any available boundary or initial conditions, which are incorporated into the total loss as supervised constraints 

By leveraging neural networks as universal function approximators, PINNs solve partial differential equations (PDEs) by embedding the underlying physical principles directly into the optimization process. A prominent application for this framework is the simulation of fluid flows governed by the nonlinear Navier-Stokes equations, which comprise temporal, convective, pressure-gradient, and diffusion terms. 
Unlike traditional computational fluid dynamics (CFD) approaches—such as Finite Difference or Finite Volume schemes—that rely on explicit spatial and temporal discretization, PINNs operate in a mesh-free manner. This characteristic makes them particularly advantageous for simulating flows within complex geometries where grid generation is traditionally a bottleneck. The defining feature of the PINN framework is its composite loss function, which simultaneously enforces adherence to the governing equations, boundary and initial conditions, and observational data, if available. This flexibility makes PINNs particularly attractive for problems involving complex geometries, inverse modeling, and data assimilation.
PINNs have been shown to produce results close to traditional CFD simulations, especially in low Reynolds number laminar conditions, where scale separation is not very large~\cite{zhao2024comprehensive}. Some examples include lid-driven cavity~\cite{kumar2025evaluation}, flow past a cylinder~\cite{ang2023physics}, flow behind airfoils~\cite{sun2023physics, kumar2024flow}, compressible flows with shocks~\cite{kumar2026robust}, hemodynamic flows~\cite{sun2020surrogate}. 

Despite these advantages, the application of PINNs to realistic fluid dynamics problems remains challenging. A central difficulty arises from the inherently multiscale nature of such systems, especially at high Reynolds numbers, where a wide spectrum of interacting length and time scales must be resolved simultaneously. Standard PINN formulations often struggle in this regime due to the spectral bias of neural networks, which favors the learning of low-frequency components while underrepresenting high-frequency features~\cite{rahaman2019spectral}. In fluid flows, these high-frequency components correspond to small-scale structures—such as thin shear layers and near-wall gradients—that are critical for accurately capturing the underlying physics. As a result, conventional PINNs tend to produce overly smooth solutions and exhibit slow or unstable convergence when applied to multiscale problems~\cite{sanderse2024scientific}. 

In addition to multiscale challenges, many practical flow configurations exhibit long-range spatial dependencies arising from the placement of boundary conditions. Unlike classical CFD solvers, which propagate information locally through iterative updates, PINNs operate as global optimizers. Consequently, in large computational domains, the effective propagation of physical information from boundaries (e.g., inlet conditions) to distant regions (e.g., outlet or recirculation zones) can be severely hindered, often leading to convergence stagnation or physically inconsistent solutions.

These challenges are further exacerbated in turbulent flow regimes. Resolving all relevant scales using the full Navier–Stokes equations requires direct numerical simulation (DNS), which is computationally prohibitive for most engineering applications~\cite{jin2021nsfnets}. As a result, reduced-order formulations based on the Reynolds-averaged Navier–Stokes (RANS) equations are commonly employed~\cite{eivazi2022physics, hanrahan2023studying,yazdani2024data}. However, the RANS framework introduces closure terms, such as the Reynolds stress tensor, which are not uniquely determined by the governing equations. In conventional CFD approaches, these terms are modeled using empirical or semi-empirical turbulence models. Within the PINN framework, however, such closure terms cannot be uniquely inferred from physics constraints alone and therefore benefit from the incorporation of auxiliary data. Although a couple of PINN studies have attempted to do it without data~\cite{harmening2024effect}, the generalizability and interpretability of learned models remain questionable.

Consequently, most existing PINN-based approaches for turbulent flows rely on extensive supervision data to compensate for both multiscale challenges and closure-related ambiguities. For instance, \citet{eivazi2022physics} solved turbulent flows over a flat plate, a NACA4412 airfoil, and a periodic hill by utilizing a comprehensive training dataset that included Reynolds-stress components on all domain boundaries. Notably, they did not employ a turbulence closure model; instead, the Reynolds-stress components were extrapolated between the boundaries. A similar approach was adopted by \citet{hanrahan2023studying} for boundary layer and periodic hill problems. However, \citet{patel2024turbulence} demonstrated that the inclusion of a turbulence closure model within the governing equations significantly improves accuracy in regions characterized by high velocity gradients, pressure gradients, and flow separation. 
%Another persistent challenge in these studies is the high sensitivity to hyperparameter selection.

The classical turbulent periodic hill problem has been investigated by several other researchers~\cite{yazdani2024data, chaurasia2024reconstruction}. Despite its complexity regarding flow separation, the periodic boundary conditions effectively eliminate the issues associated with long-range dependency. Some investigators have also attempted to use PINN for predicting more complex flows such as Backward-facing Step (BFS; ~\cite{nadge2014high}). This configuration is characterized by several complex phenomena, including adverse pressure gradients, flow separation induced by sudden geometric expansion, the formation and evolution of primary and secondary recirculation bubbles, boundary-layer development downstream of the step, and turbulent dissipation at higher Reynolds numbers. The computational challenge of the BFS problem stems from the large domain, which requires physical information to propagate from the inlet to the distant outlet while negotiating the geometric step.  \citet{pioch2023turbulence} attempted to solve the turbulent BFS flow by utilizing data along multiple streamwise locations—a strategy that effectively bypasses the challenge of long-range information propagation. A comparable approach was taken by \citet{chaurasia2024reconstruction} for a curved backward-facing step flow.

In forward problems, the excessive requirement of data at specific locations often diminishes the principal advantage of PINNs—namely, reduced dependence on high-fidelity data—since such datasets are typically available only through computationally expensive DNS or large-eddy simulation (LES), or through sophisticated experimental techniques such as particle image velocimetry (PIV).  In fact, if enough data is available, operator-based learning frameworks—such as the Neural Operator \cite{kovachki2023neural}, Fourier Neural Operator (FNO) \cite{li2023fourier}, Graph Neural Operator (GNO) \cite{li2020neural}, and Transolver++ \cite{luo2025transolver++}—may offer superior computational efficiency. In the PINN approach, if the architecture is capable of adequately representing multiscale features, the dependence on dense supervision can be significantly alleviated. In such scenarios, even sparse measurements, such as pointwise probe measurements or limited PIV fields, may enable the reconstruction of high-resolution flow fields across the domain. This can be interpreted as a form of physics-guided super-resolution, wherein limited data are leveraged in conjunction with governing equations to recover detailed solutions.

Therefore, to render PINNs both tractable and viable for practical engineering flow problems with minimal or no data, a framework must robustly address the following three pillars: 1) Strong Nonlinearity: The capacity to resolve steep gradients, non-periodic, or transient physical events without numerical instability. 2) Multiscale Resolution: The ability to simultaneously capture a broad spectrum of relevant physical scales, from large-scale eddies to dissipative structures. 3) Long-Range Dependency: An architectural mechanism to effectively propagate physical information across expansive domains with distally placed boundaries.

To address nonlinearity and stiffness, \citet{McClenny2023} and \citet{anagnostopoulos2024residual} proposed the Residual-based adaptive weighting approaches (RBA-PINN) approach, in which local adaptive trainable weights are optimized via a min-max formulation. In this framework, the total loss is minimized with respect to the network parameters while simultaneously maximizing the weights assigned to local residual points. This mechanism forces the model to prioritize high-residual areas—typically where nonlinearities are strongest—effectively compelling the network to resolve sharp physical features that a standard global optimizer might otherwise smooth over.

To improve efficiency and accuracy in multiscale problems, \citet{dwivedi2021distributed} introduced the Distributed Learning Machine (DLM). This framework partitions the domain into non-overlapping subdomains and enforces interface-flux continuity via regularization terms. While promising for large-scale PDEs, this approach increases computational complexity due to the separate formulations for PINN-based losses and interface regularizers within the global loss. Subsequently, \citet{moseley2023finite} proposed the Finite Basis PINN (FB-PINN), which utilizes overlapping subdomains defined by window functions. In this architecture, each subdomain is represented by an independent neural network with input–output normalization to enhance local convergence. However, the use of individual local loss functions can lead to inconsistencies at the interfaces. For complex, sensitive flows—particularly those with long-range dependencies—even minor inconsistencies in interfacial information exchange can result in unphysical behavior or divergence. Furthermore, while input–output normalization is often beneficial, it may not be practical for all fluid dynamics problems, where relevant scales for normalization may not be present.

While the above-mentioned approaches have demonstrated improvements in convergence and accuracy, they introduce additional complexities, including interface inconsistencies, increased computational overhead, and sensitivity to hyperparameter selection. Moreover, many of these methods do not explicitly address long-range dependency issues in large computational domains. To overcome these challenges, we propose a Domain-Decomposed and Shifted Physics-Informed Neural Network (DDS-PINN) framework, designed to simultaneously address nonlinearity, multiscale resolution, and long-range dependency. The key idea is to decompose the computational domain into overlapping subdomains, within which localized neural networks operate on shifted input coordinates.

By shifting input coordinates toward zero within localized subdomains, DDS-PINN effectively mitigates vanishing gradient issues, thereby accelerating convergence and enhancing the network's ability to capture local flow features. This localized shifting operation also renders global input–output normalization unnecessary, avoiding the practical limitations often encountered in traditional fluid dynamics scaling. Furthermore, unlike FB-PINN, DDS-PINN utilizes a single global loss function evaluated across the decomposed domain, which ensures strict surface consistency and prevents interfacial discontinuities.  To further enhance convergence in regions with strong nonlinearities, a residual-based attention (RBA) training strategy is incorporated into the DDS-PINN framework.

The application of DDS-PINN is demonstrated across three benchmark problems of increasing multiscale complexity and long-range dependency: 1) a multiscale ordinary differential equation (ODE), 2) the unsteady Burgers’ equation, and 3) laminar flow over a flat plate at $Re = 500$. Subsequently, the framework is employed to predict computationally challenging and complex incompressible BFS flow, described earlier. 
For low Reynolds numbers under laminar conditions, DDS-PINN is employed in a data-free manner. For turbulent regimes, the framework utilizes the Reynolds-Averaged Navier-Stokes (RANS) equations as physical constraints, supplemented by a sparse supervision dataset. However, in contrast to earlier studies that use data at several streamwise locations or at all boundaries to get around the long-range dependency issue, our framework is specifically designed to enhance resolution from sparse, localized sources, such as low-resolution PIV or coarse simulations. By resolving the long-range dependencies inherent in the BFS geometry without dense spatial scaffolding, our method aligns more closely with super-resolution tasks, where physics-informed constraints are leveraged to reconstruct high-fidelity flow fields from minimal localized information.

\section{Methodology}
\label{sec:dds}

In conventional PINNs, the neural network approximates the solution \( u(\mathbf{x}) \) globally over \(\Omega\). The training dynamics are governed by the Neural Tangent Kernel (NTK\cite{jacot2018neural}) \( K(\mathbf{x}, \mathbf{x}') \), which describes the network’s learning behavior in the infinite-width limit. The NTK’s eigenvalue spectrum decays rapidly for high-frequency modes, causing the network to prioritize low-frequency components during gradient descent~\cite{khodakarami2026spectral}. Formally, the eigenvalue problem:  
\begin{equation}
    \int_\Omega K(\mathbf{x}, \mathbf{x}') \psi_k(\mathbf{x}') d\mathbf{x}' = \lambda_k \psi_k(\mathbf{x}),
\end{equation},
reveals that eigenfunctions \(\psi_k(\mathbf{x})\) associated with small eigenvalues \(\lambda_k\) (high frequencies) are learned more slowly than those with large \(\lambda_k\) (low frequencies). This results in poor resolution of high-frequency features. Because of these reasons, conventional PINNs often struggle with stiff or multiscale problems because their global network architecture tends to homogenize localized features. In this work, we employ the Domain-Decomposed and Shifted Physics-Informed Neural Network (DDS-PINN) framework to solve complex fluid-dynamical problems.

\subsection{DDS-PINN Framework}
The DDS-PINN architecture is designed to approximate solutions $u(\mathbf{x}) \in \mathbb{R}^n$ of partial differential equations (PDEs) over intricate overlapping spatiotemporal domains $\Omega \subset \mathbb{R}^m$. 
% To overcome this, DDS-PINN decomposes $\Omega$ into $N$ overlapping subdomains $\{\Omega_i\}$. 
Each subdomain is equipped with a dedicated neural network $\mathcal{N}_i$ and a smooth window function $w_i(\mathbf{x})$ that blends local solutions into a globally consistent output. The global solution is synthesized as:
\begin{equation}
    u(\mathbf{x}) = \sum_{i=1}^N w_i(\mathbf{x}) \cdot \mathcal{N}_i\left( \mathbf{x} - \mathbf{S}_i\right)
\end{equation}
where, $\mathbf{S}_i \in \mathbb{R}^m$ represents a shift vector that centers the input coordinates $\mathbf{x}$ within the respective subdomain $\Omega_i$.

\begin{figure}
    \centering
\subfloat[\centering DDS-PINN Framework]
        {{\includegraphics[width=0.57\textwidth]{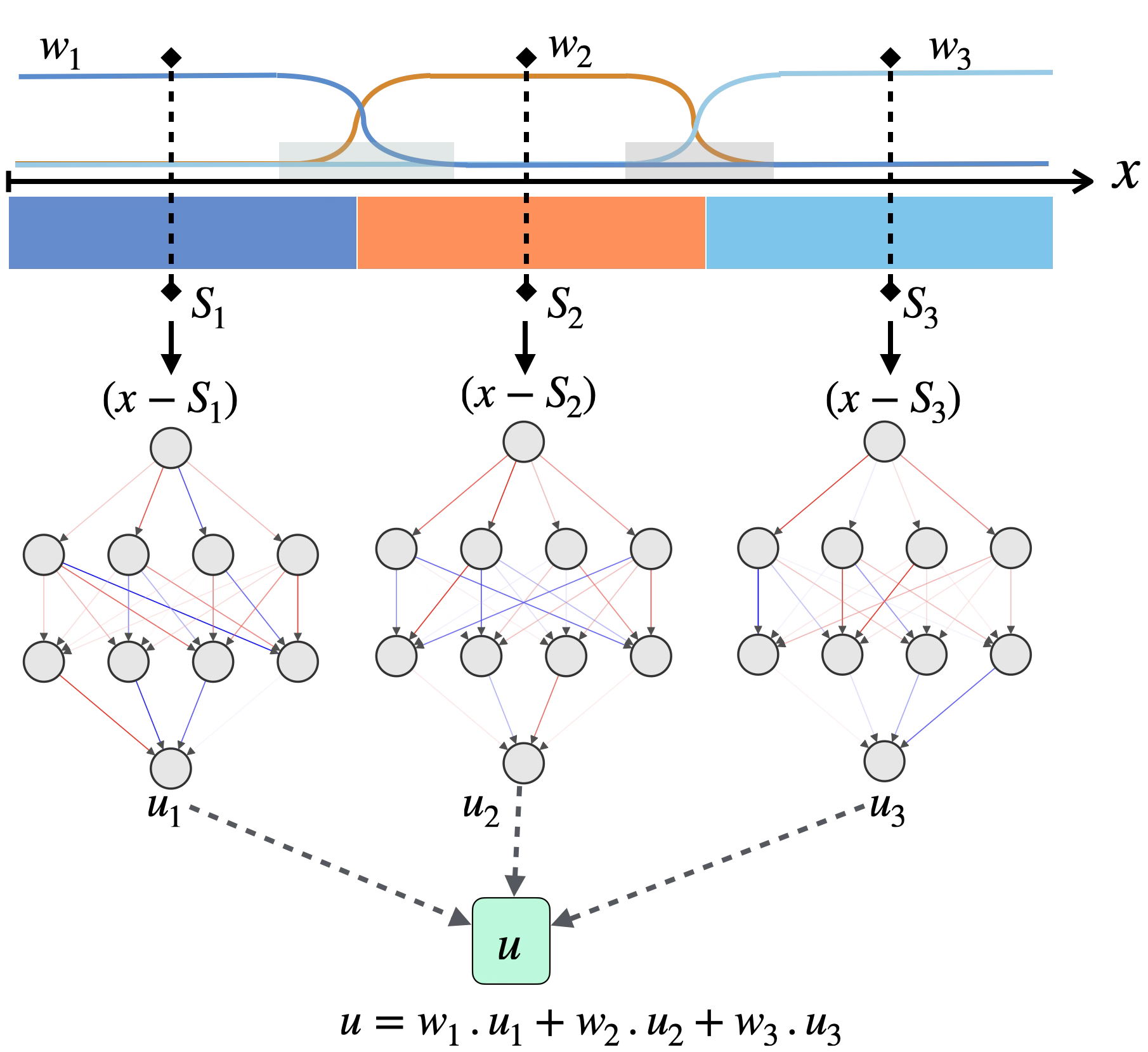} }}
         \subfloat[\centering Sub-Domain prediction]
        {{\includegraphics[width=0.43\textwidth]{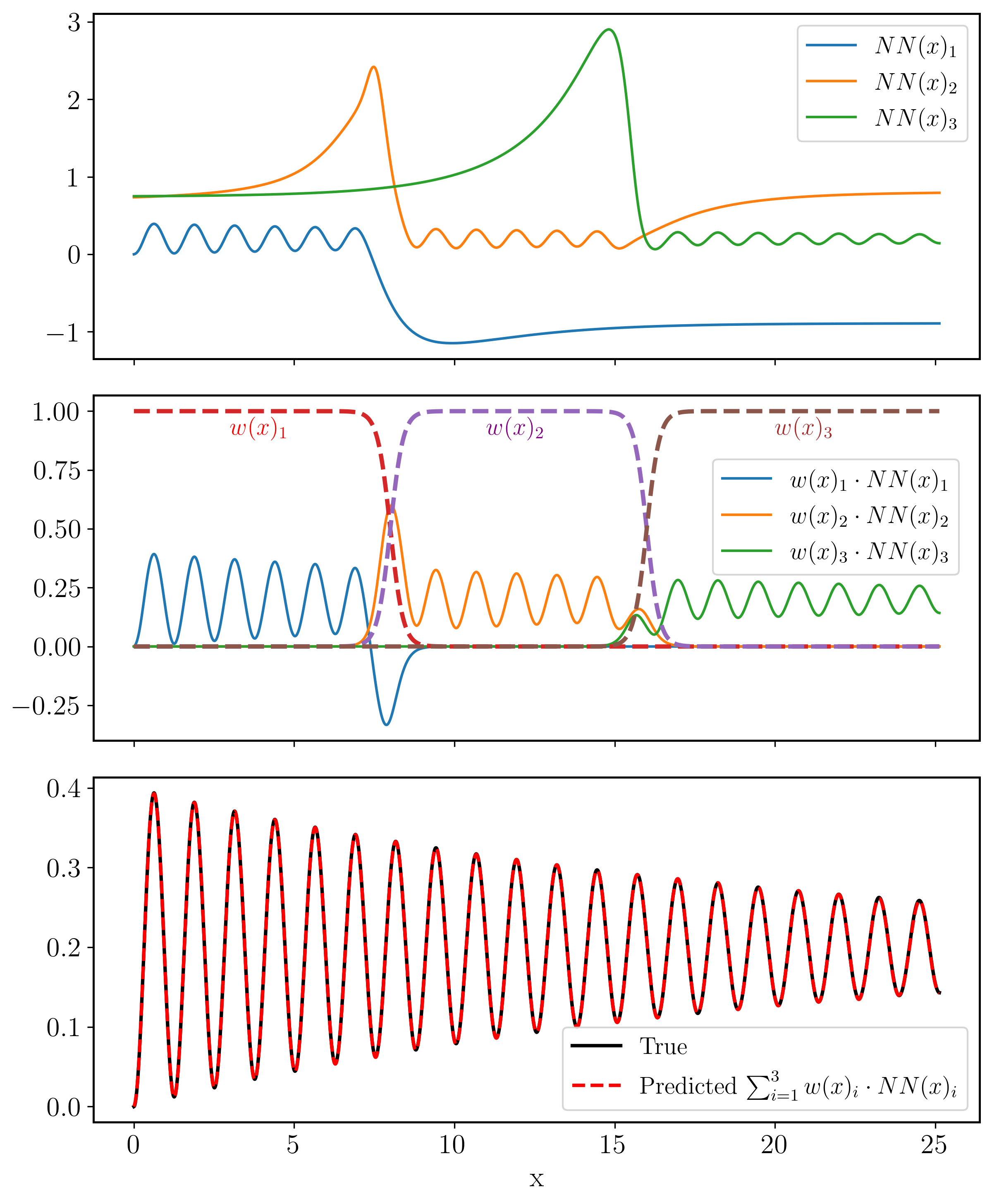} }}
        \caption{DDS-PINN framework using 3 subdomains. (a) Architecture, (b) Predictions for multiscale ODE~\eqref{eqref:multiscaleode} }
            \label{fig:network}
\end{figure}        

Figure~\ref{fig:network}(a) illustrates a typical domain decomposition and shifting implementation for three subdomains with their own sub-networks. Each sub-network receives inputs shifted by vectors $(S_1, S_2, S_3)$, corresponding to the approximate centers of their respective subdomains. The outputs $(u_1, u_2, u_3)$ of these sub-networks are multiplied by non-trainable window functions $(w_1, w_2, w_3)$. These window functions are constructed as tensor products of one-dimensional sigmoid-based kernels in each spatial direction. For a given subdomain, the window smoothly activates within the prescribed bounds and decays to zero outside using sigmoid transition functions with specified scaling parameters. Specifically, the window functions $w_i(\mathbf{x})$ satisfy a partition of unity ($\sum_{i=1}^N w_i(\mathbf{x}) \approx 1$), ensuring smooth transitions between subdomains and eliminating artificial discontinuities.

The DDS-PINN architecture avoids the "low-frequency dominance"  issue by localizing the learning task to smaller subdomains \(\Omega_i\), where the effective frequency content of the solution is reduced. Each subnetwork \(\mathcal{N}_i\) operates on shifted coordinates \(\mathbf{x} - \mathbf{S}_i\), effectively recentering the subdomain \(\Omega_i\) to the origin. Suppose the original solution \(u(\mathbf{x})\) contains high-frequency components in a region \(\Omega_i\). After shifting, the local solution \(u_i(\mathbf{x}') = u(\mathbf{x}' + \mathbf{S}_i)\) within \(\Omega_i\) is represented in a coordinate system where spatial variations are rescaled. For example, a high-frequency mode \( \sin(kx) \) in \(\Omega\) becomes \( \sin(k(x' + S_i)) \), which retains the same frequency \(k\) but is learned within a localized region. Crucially, the effective bandwidth required to represent \(u_i(\mathbf{x}')\) over the smaller \(\Omega_i\) is reduced because the subnetwork only needs to resolve frequencies relevant to the subdomain, not the entire \(\Omega\).  

DDS approach thus enables localized training while strictly preserving global physical constraints, making the framework particularly effective for problems characterized by sharp gradients, multiscale dynamics, long-range dependencies or irregular geometries.
Notably, unlike the FB-PINN approach~\cite{moseley2023finite}, which normalizes both inputs and outputs—potentially distorting spectral bias through output scaling—our method preserves inherent spectral characteristics by only shifting inputs to subdomain centers and avoiding output normalization. 
The shifting operation also improves numerical conditioning by aligning inputs with the subnetwork’s specific region of focus, thereby mitigating gradient pathologies common in large-scale domains.

\subsubsection{Subdomain Network}
Each subnetwork \(\mathcal{N}_i\) employs a specialized architecture to enhance feature extraction and gradient propagation. For a $n$-D input \(\mathbf{x} = (x_1, x_2, ..., x_n)\), the network first processes coordinates through dual encoder branches:  
\[
\mathbf{U} = E_U(\mathbf{x}) \in \mathbb{R}^d, \quad \mathbf{V} = E_V(\mathbf{x}) \in \mathbb{R}^d,
\]  
where \(E_U\) and \(E_V\) map \(\mathbf{x}\) into latent feature spaces \(\mathbb{R}^d\). These encoders capture complementary spatial or spectral representations of the input. The features are then fused through a sequence of nonlinear layers with residual mixing, which dynamically combines encoder outputs using gating mechanisms. For example, at layer \(l\):  
\[
\mathbf{z}_1 = \tanh(\mathbf{W}_1 \mathbf{z}_0 + \mathbf{b}_1),
\]  
\[
\mathbf{z}_2 = (1 - \mathbf{z}_1) \odot \mathbf{U} + \mathbf{z}_1 \odot \mathbf{V},
\]  
where \(\odot\) denotes element-wise multiplication. Here, \(\mathbf{z}_1\) acts as a learned gate, interpolating between the encoder features \(\mathbf{U}\) and \(\mathbf{V}\) to adaptively emphasize different input characteristics. Subsequent layers refine these mixed features:  
\[
\mathbf{z}_3 = \tanh(\mathbf{W}_2 \mathbf{z}_2 + \mathbf{b}_2), \quad \ldots, \quad \mathbf{z}_L = \tanh(\mathbf{W}_L \mathbf{z}_{L-1} + \mathbf{b}_L).
\]  
This architecture promotes stable training by mitigating vanishing gradients and enabling hierarchical feature learning, crucial for resolving high-frequency or discontinuous phenomena.

\subsubsection{Loss function}
Unlike FB-PINN~\cite{moseley2023finite}, which utilizes an aggregate of local losses across subdomains, the training process in DDS-PINN enforces physical constraints via a global weighted loss function:
\begin{equation}
    \mathcal{L}_{\text{total}} = \lambda_{\text{IC}} \mathcal{L}_{\text{IC}} + \lambda_{\text{BC}} \mathcal{L}_{\text{BC}} + \lambda_{\text{PDE}} \mathcal{L}_{\text{PDE}} + \lambda_{\text{Data}} \mathcal{L}_{\text{Data}}
\end{equation}
where $\mathcal{L}_{\text{IC}}$ and $\mathcal{L}_{\text{BC}}$ penalize deviations from the initial and boundary conditions using the mean squared error (e.g., $\mathcal{L}_{IC} = \frac{1}{N_{IC}} \sum_{i=1}^{N_{IC}} | u_\theta(x_i^{IC}) - u_i^{IC} |^2$). The term $\mathcal{L}_{\text{PDE}}$ computes the PDE residual:
\begin{equation}
    \frac{1}{N_{PDE}}\sum_{i=1}^{N_{PDE}}|\lambda_{i}\mathcal{F}(u, \nabla u, \dots)_i|^2
\end{equation}
at collocation points, where $\lambda_{i}$ represents the local residual parameters. For improving accuracy at high-gradient regions, these local parameters are obtained using RBA~\cite{anagnostopoulos2024residual}. Additionally, $\mathcal{L}_{\text{Data}}$ incorporates experimental or synthetic data into the training objective. The weights $\{\lambda_{\text{IC}}, \lambda_{\text{BC}}, \dots\}$ are employed to balance these competing objectives. This global loss formulation ensures the elimination of interface inconsistencies, distinguishing it from DLM~\cite{dwivedi2021distributed}, which relies on interface regularizers alongside local PINN losses to maintain consistency across domains.

The networks are trained to minimize the total loss using the Adam optimizer with single-precision (FP32) parameters, inputs, and outputs for all cases. The number of training epochs is case-dependent and is detailed in the respective sections. The implementation is carried out in PyTorch $(v2.6.0+cu124)$ and executed on an NVIDIA H100 GPU.

\subsection{Ablation Study and Comparisons with Existing Approaches}
We first show an ablation study considering the stiff first-order linear ODE, given as: 
\begin{equation}\label{eqref:multiscaleode}
\frac{du}{dx} = e^{-x/20} \sin(5x), \quad  u(0) = 0 
\end{equation}

The analytical solution, given below, exhibits decaying behavior with high-frequency components. 
\begin{equation}
u(x) = \frac{2000 - 20\left( \sin(5x) + 100 \cos(5x) \right) e^{-x/20}}{10001}
\end{equation}

\begin{figure}
\centering
    \subfloat[\centering Vanilla PINN]
        {{\includegraphics[width=0.31\textwidth]{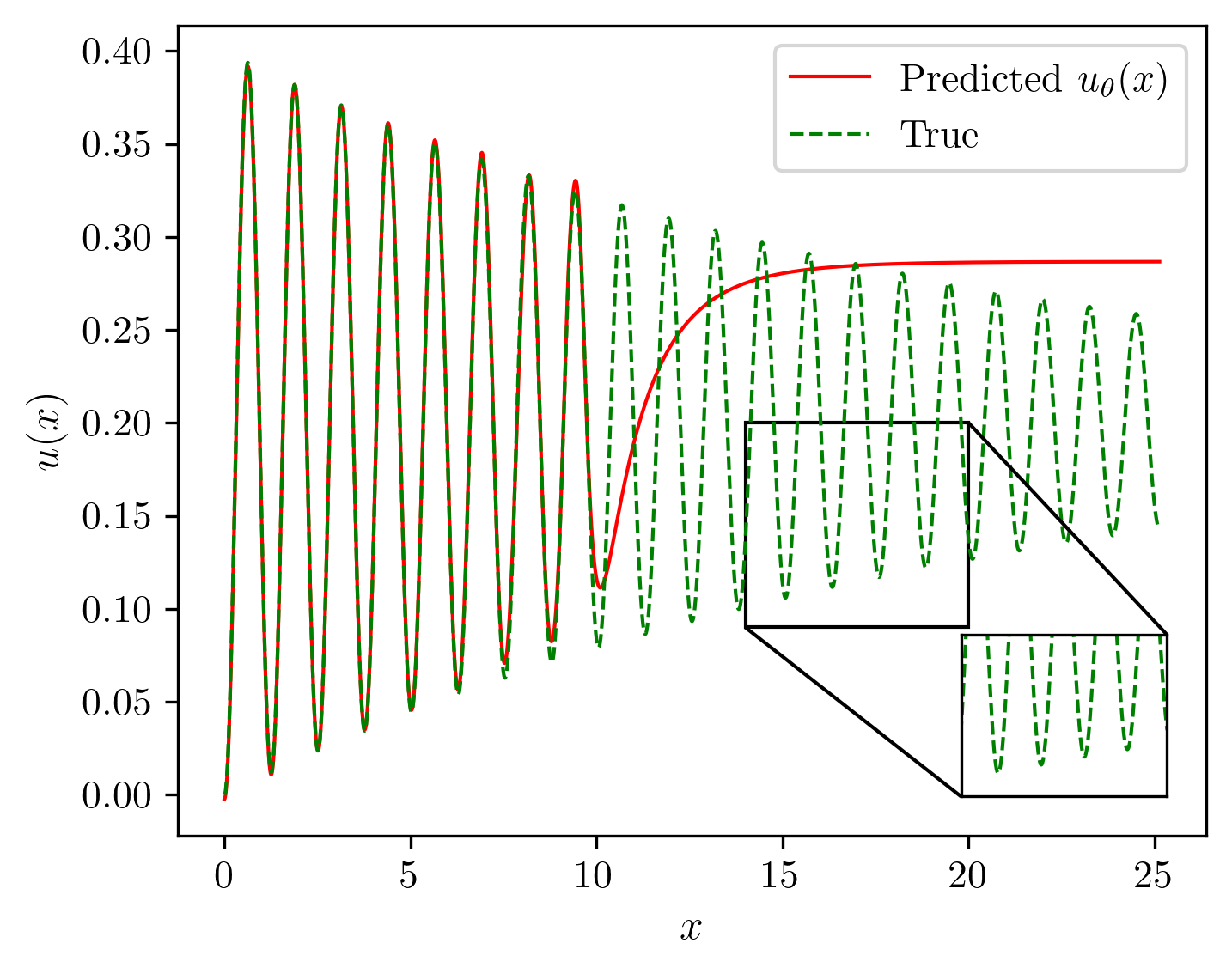} }}
        \subfloat[\centering Using domain-decomposition (DD)]
        {{\includegraphics[width=0.31\textwidth]{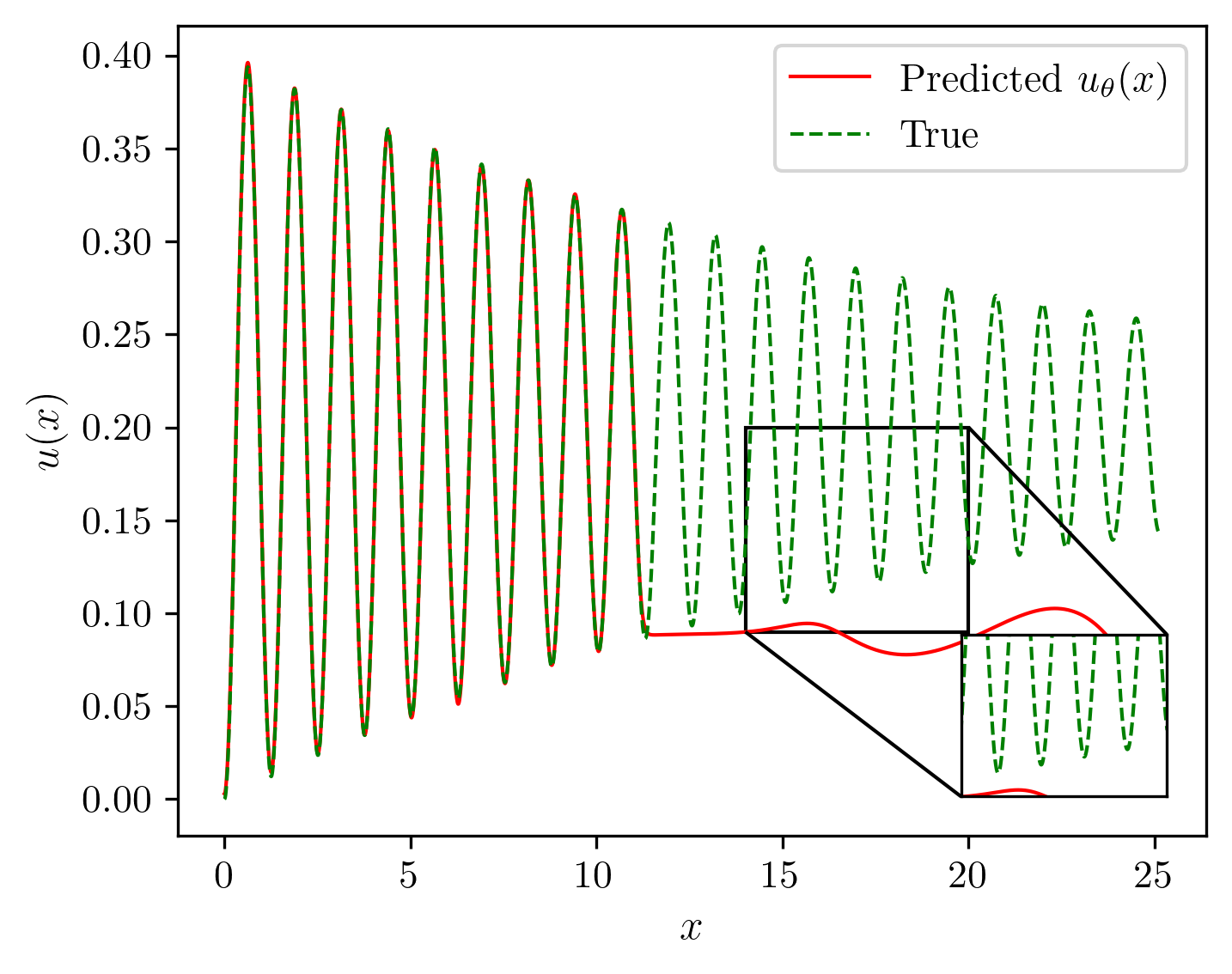} }}
        \subfloat[\centering Using domain-decomposition and shifting (DDS) but without RBA.]
        {{\includegraphics[width=0.31\textwidth]{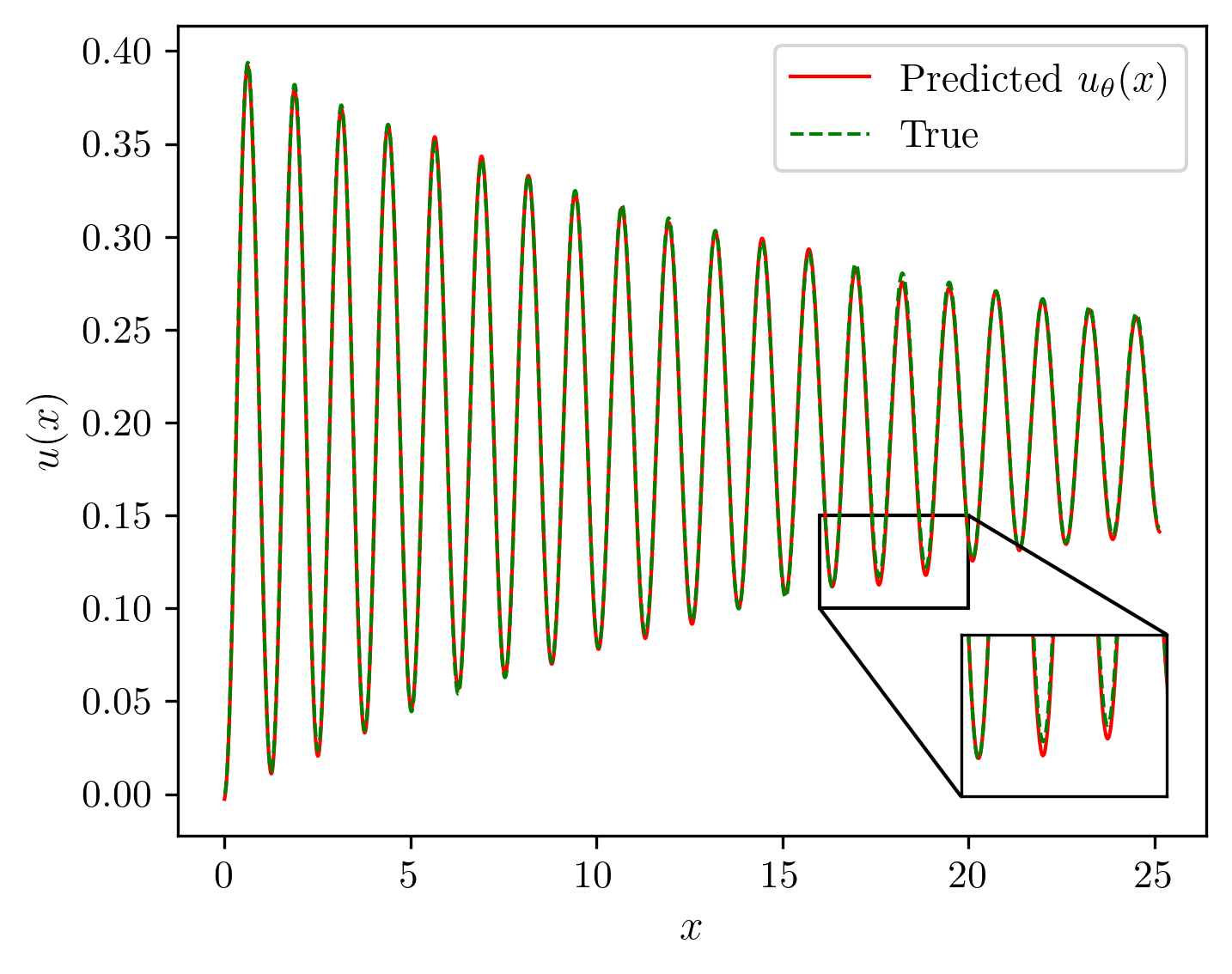} }} \\ 
               \subfloat[\centering FB-PINN (after $100{,}000$ epochs)]
        {{\includegraphics[width=0.31\textwidth]{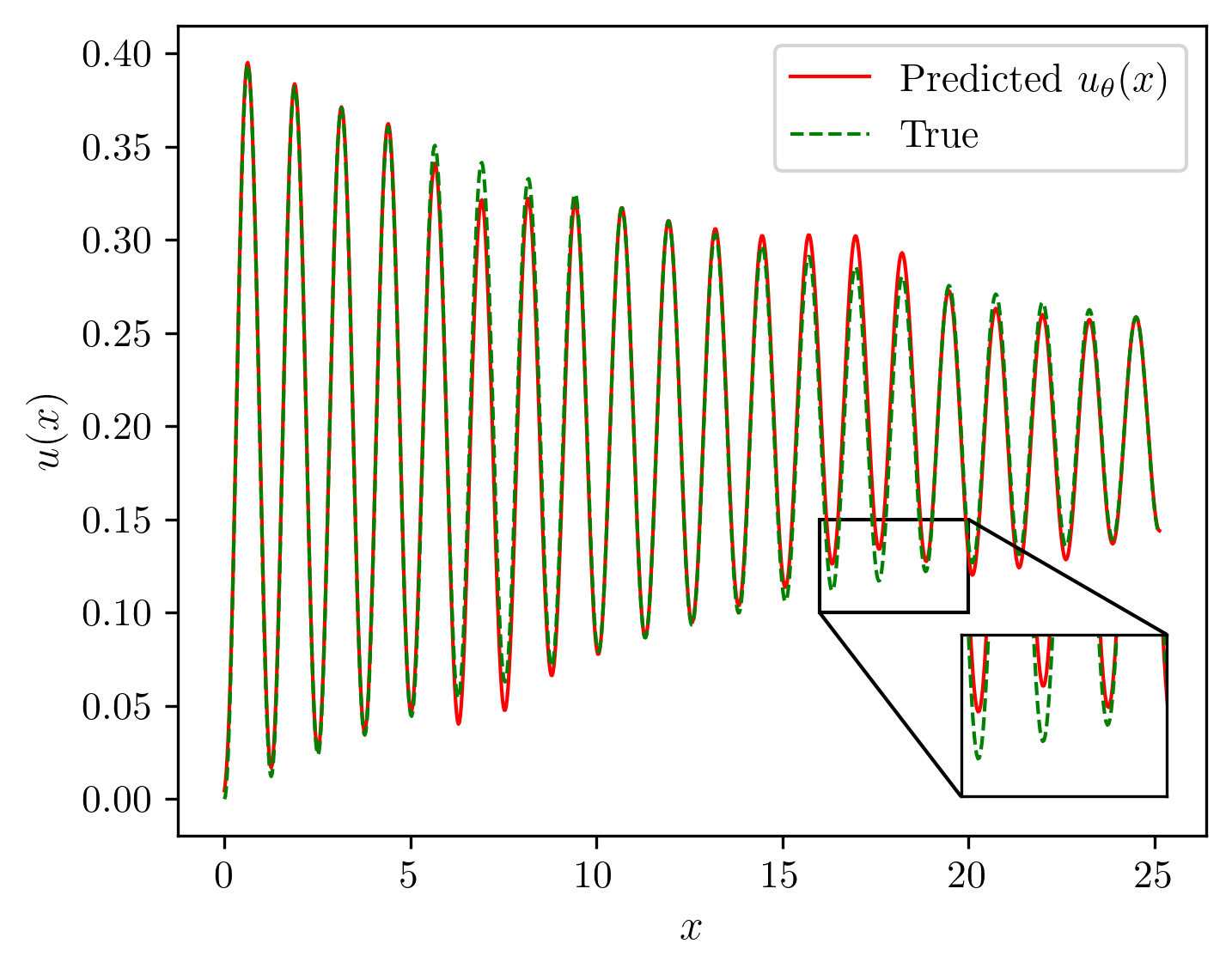} }}  
        \subfloat[\centering  RBA-PINN]
        {{\includegraphics[width=0.31\textwidth]{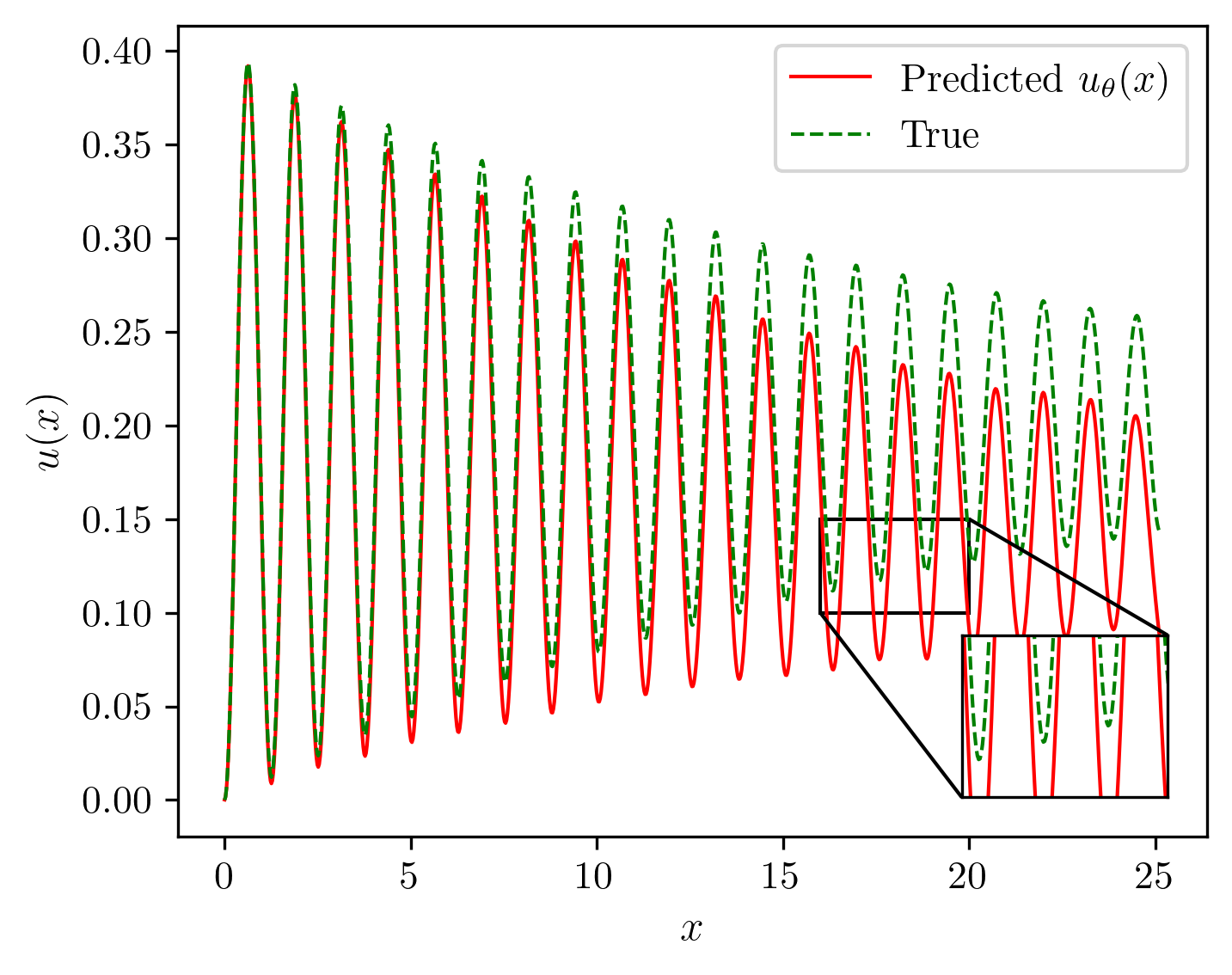} }}
                \subfloat[\centering Using DDS with RBA]
        {{\includegraphics[width=0.31\textwidth]{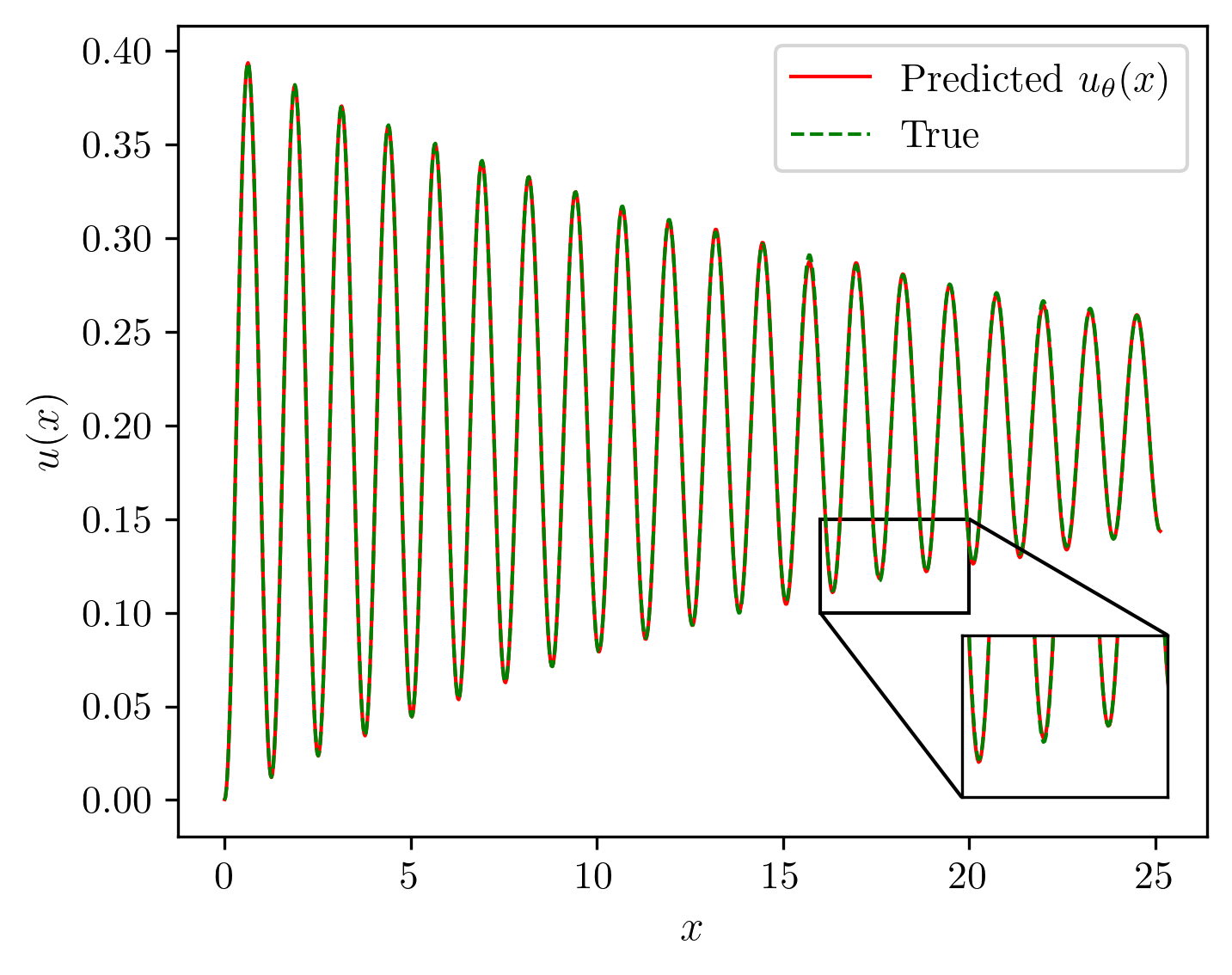} }}
        \\
        \subfloat[\centering Total loss vs epoch comparison.]
        {{\includegraphics[width=0.6\textwidth]{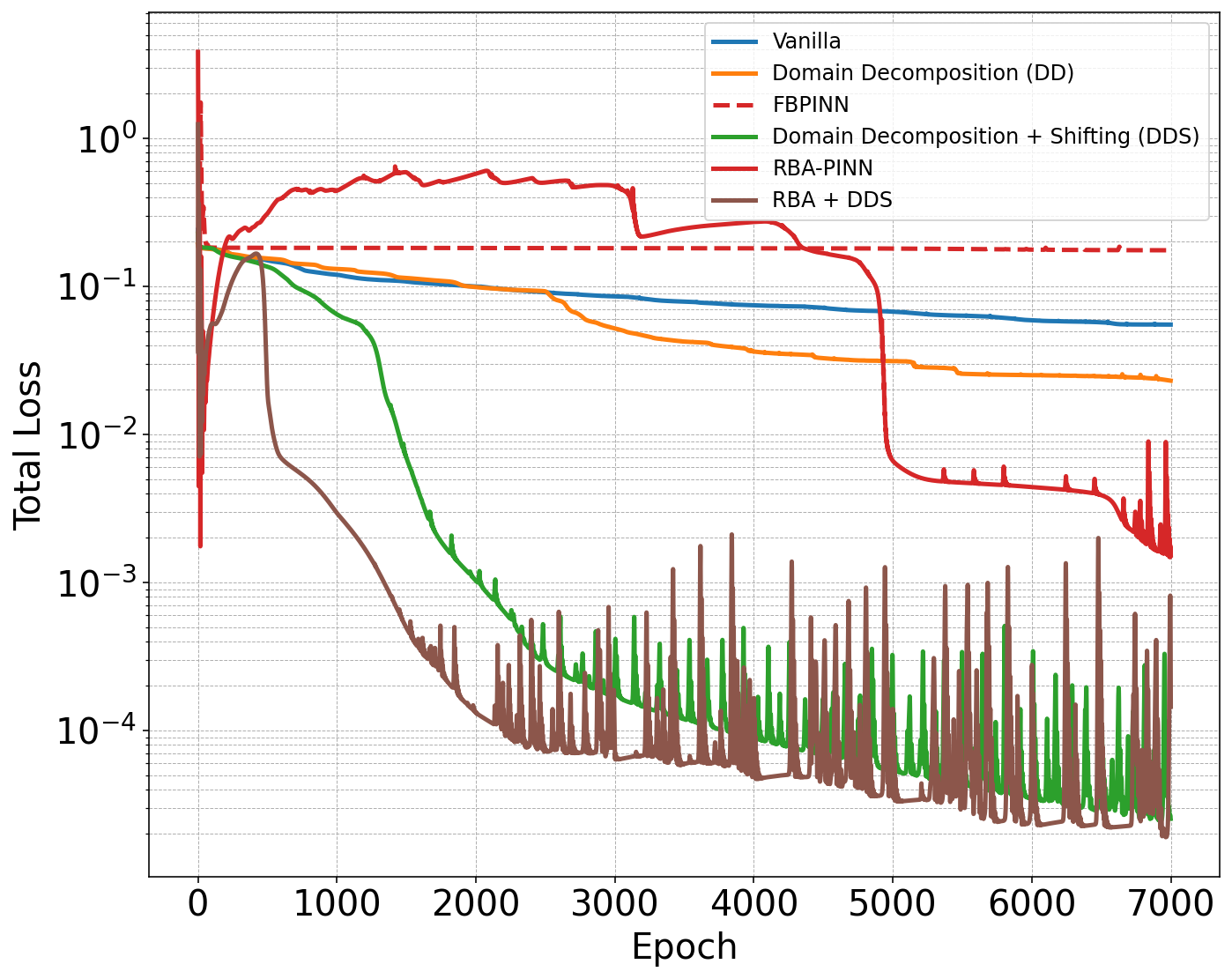} }}
    \caption{Solutions of multiscale ODE problem using different PINN approaches}
    \label{fig:1d}
\end{figure}

This problem is solved on the domain $[0, 8\pi]$ using three physics-informed learning strategies with increasing complexity: 1) vanilla PINN implemented on a single domain; 2) domain-decomposed PINN (without shifting) with three subdomains ($[0, 8]$, $[8, 16]$ and $[16, 8\pi]$) and 3) domain-decomposed PINN with three shifted subdomains ($S_1, S_2, S_3$)=($4.0, 12.0, 20.0$). Along with the ablation study, results are also compared with existing approaches in the literature: 1) FB-PINN and  2) RBA-PINN implemented on a single domain with residual-based adaptive (RBA); and 3)  RBA-PINN implemented employing three shifted subdomains. The network information for these approaches is provided in Table~\ref{tab:ablation}. The table also includes the total loss after 7000 epochs.  

\begin{table}
    \centering
\caption{\centering Ablation study for a first-order linear ODE. \newline NN: neural network architecture \newline HL: Number of hidden layers (\# subdomains $\cdot$ (layers $\times$ [neurons])) \newline TL: Total loss after 7{,}000 epochs.}
\label{tab:ablation}
    \begin{tabular}{|c|c|c|c|c|c|c|}\hline
         NN&  Vanilla&  DD&  DDS &  FB-PINN&  RBA& DDS+RBA\\\hline
         HL&  $1\cdot(3\times[32])$&  $3\cdot(3\times[16])$&  $3\cdot(3\times[16])$&  $3\cdot(3\times[16])$&  $1\cdot(3\times[32])$& $3\cdot(3\times[16])$\\\hline
         TL&  0.0522&  0.024&  $2.53\times 10^{-5}$&  0.1778&  $1.5\times 10^{-3}$& $1.90\times 10^{-5}$\\ \hline
    \end{tabular}
\end{table}

As shown in Figure~\ref{fig:1d}a, the default vanilla PINN fails to accurately predict the solution beyond $x > 10$. Introducing domain decomposition (without shifting) only slightly improves performance, as illustrated in Figure~\ref{fig:1d}b; however, noticeable deviations from the analytical solution persist. The loss landscape for all approaches is shown in Figure~\ref{fig:1d}g up to 7000 epochs, and we note a slight improvement in convergence for the decomposed PINN compared to the vanilla approach. Note that training for further epochs may improve the performance, but will increase the computational cost. 

A key challenge in this setting arises from the spatial extent of the input domain, which lies in a numerical range where many commonly used activation functions are prone to vanishing gradients. This behavior significantly hinders training and convergence. Consequently, appropriate input preprocessing—specifically, shifting the inputs to recenter each subdomain—plays a crucial role in alleviating gradient decay and improving learning efficiency. This improvement is evident in the DDS framework results (Figure~\ref{fig:1d}c), where predictions closely match the analytical solution and achieve convergence on the order of $O(10^{-5})$ (see Table~\ref{tab:ablation} and Figure~\ref{fig:1d}g). To further illustrate the contributions of the individual subnetworks, the top panel of Figure~\ref{fig:network}b displays the corresponding subdomain-wise predictions. The middle panel shows the window functions $w_i(\mathbf{x})$, which transition smoothly from 1 to 0 from their native subdomain to the adjacent ones. Finally, the bottom panel presents the reconstructed global prediction.

Figure~\ref{fig:1d}d presents the results obtained using FB-PINN. Although this approach demonstrates the capability to handle the stiff ODE, significant amplitude errors persist. Furthermore, the results shown are obtained after $100,000$ epochs, as the convergence rate is relatively slow—the loss remains high even after $7,000$ epochs, as detailed in Table~\ref{tab:ablation} and Figure~\ref{fig:1d}g. Additionally, sensitivity studies using 20, 40, and 60 subdomains—with single hidden layers of 16 and 32 neurons—reveal that the loss oscillates within the $O(10^{-3})$ to $O(10^{-4})$ range across all configurations. In contrast, the DDS framework achieves a significantly lower convergence threshold of $O(10^{-5})$. The improved performance of DDS over FB-PINNs may be attributed to its globally coupled domain-decomposed architecture, which preserves cross-subdomain gradient interactions and leads to more efficient optimization and faster convergence.

Figure~\ref{fig:1d}e presents the results obtained using the Residual-Based Attention (RBA) method after 7,000 epochs. While RBA outperforms both the vanilla and basic domain-decomposed networks, its accuracy remains inferior to that of the DDS approach. This improvement in RBA can be attributed to the incorporation of positional encoding for input features and the adaptive weighting of residual points. Figure~\ref{fig:1d}f illustrates the integrated DDS+RBA approach, which further enhances prediction accuracy. It also achieves superior and further convergence compared to DDS alone, as shown in Figure~\ref{fig:1d}g. Consequently, the DDS with RBA approach is adopted as the default configuration for the subsequent fluid-dynamics problems presented in this study. For convenience, this approach is referred to as `DDS-PINN'.

\section{Benchmark Problems}
\subsection{1D Burgers' Equation}
We employ DDS-PINN for predictions of the 1D Burgers' equation—a nonlinear, diffusive system, given in the conservative form as:  
\begin{equation}
    \frac{\partial u}{\partial t} + \frac{\partial (\frac{1}{2}u^2)}{\partial x} = \nu \frac{\partial^2 u}{\partial x^2}
\end{equation}
This system serves as a classical benchmark for numerical methods, as the convective nonlinearity leads to steepening gradients and the eventual formation of shock-like discontinuities. The Burgers' equation is solved over the domain $x \in [0,1]$ and $t \in [0,1]$, with a viscosity of $\nu = 0.01$ and the initial condition $u(x,0) = \sin(2\pi x)$. The DDS-PINN approach is used in a completely data-free setting with the temporal domain decomposed into three subdomains.

\begin{figure}
\centering
    \subfloat[\centering Domain-decomposed solutions]
  {{\includegraphics[width=1.0\linewidth]{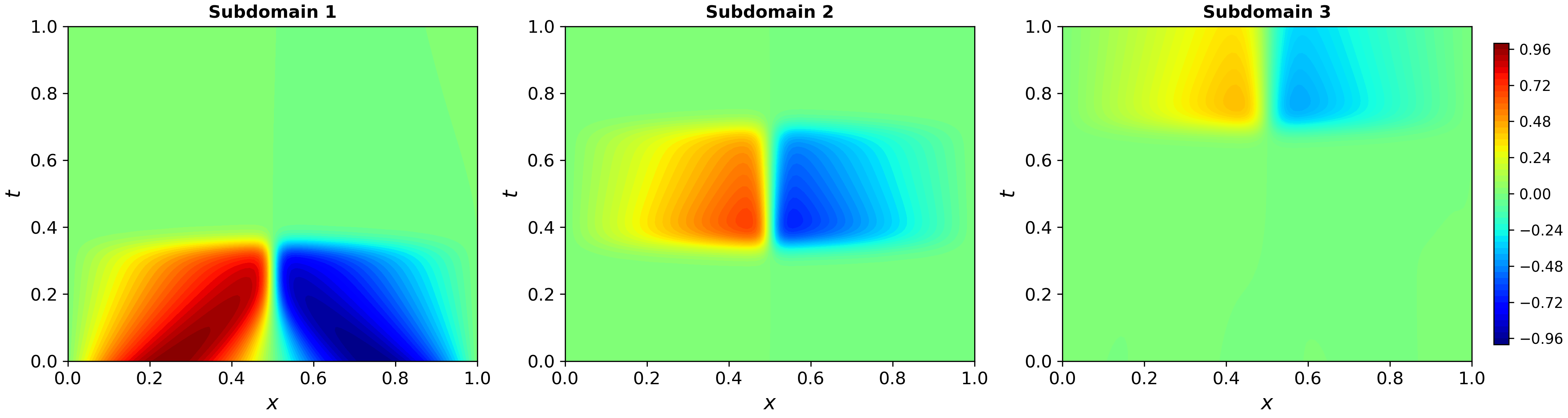}}} \\
     \subfloat[\centering DDS-PINN Prediction]
        {{\includegraphics[width=0.44\textwidth]{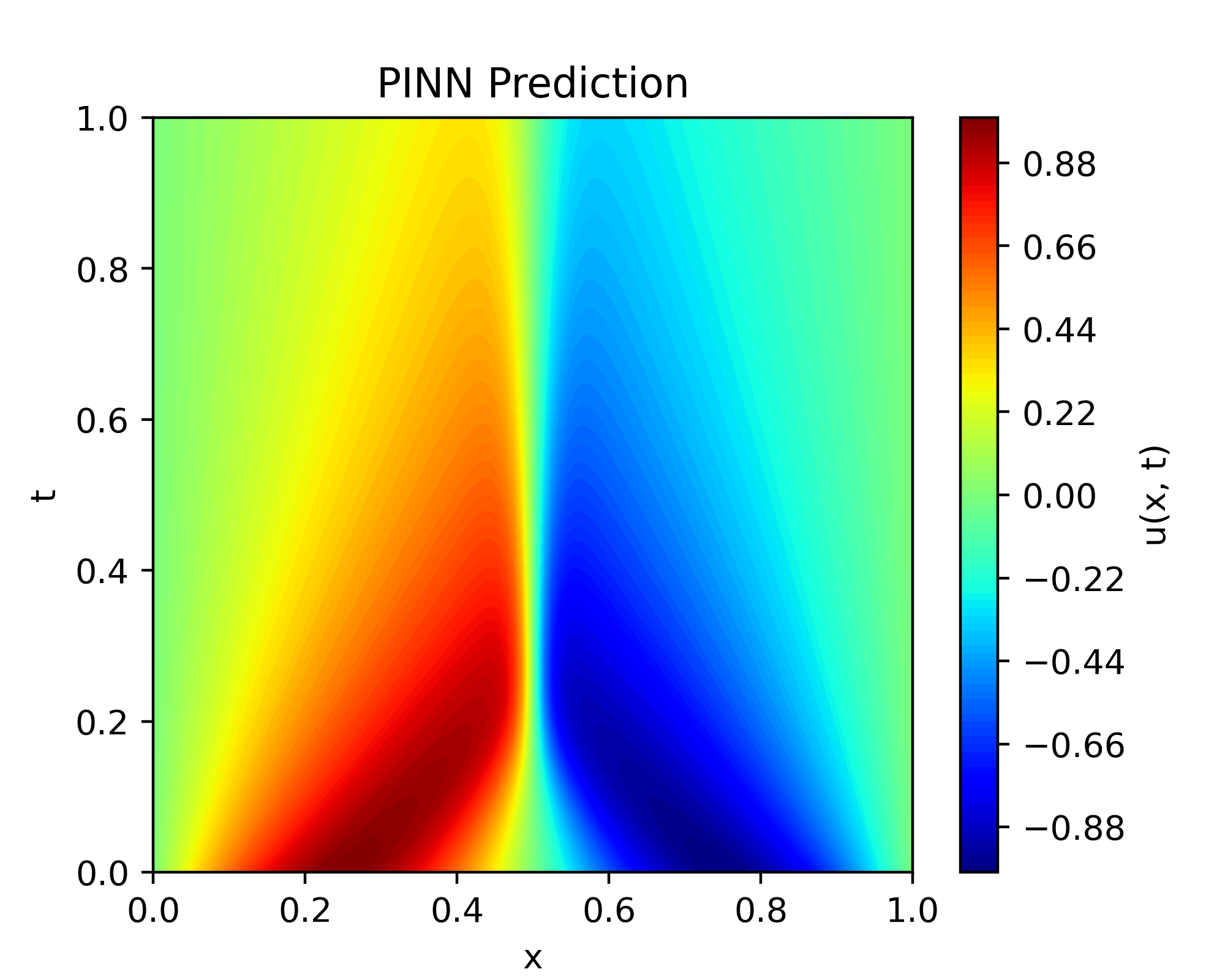} }}
        \subfloat[\centering Comparison with CFD (solid black line) at different times]
        {{\includegraphics[width=0.49\textwidth]{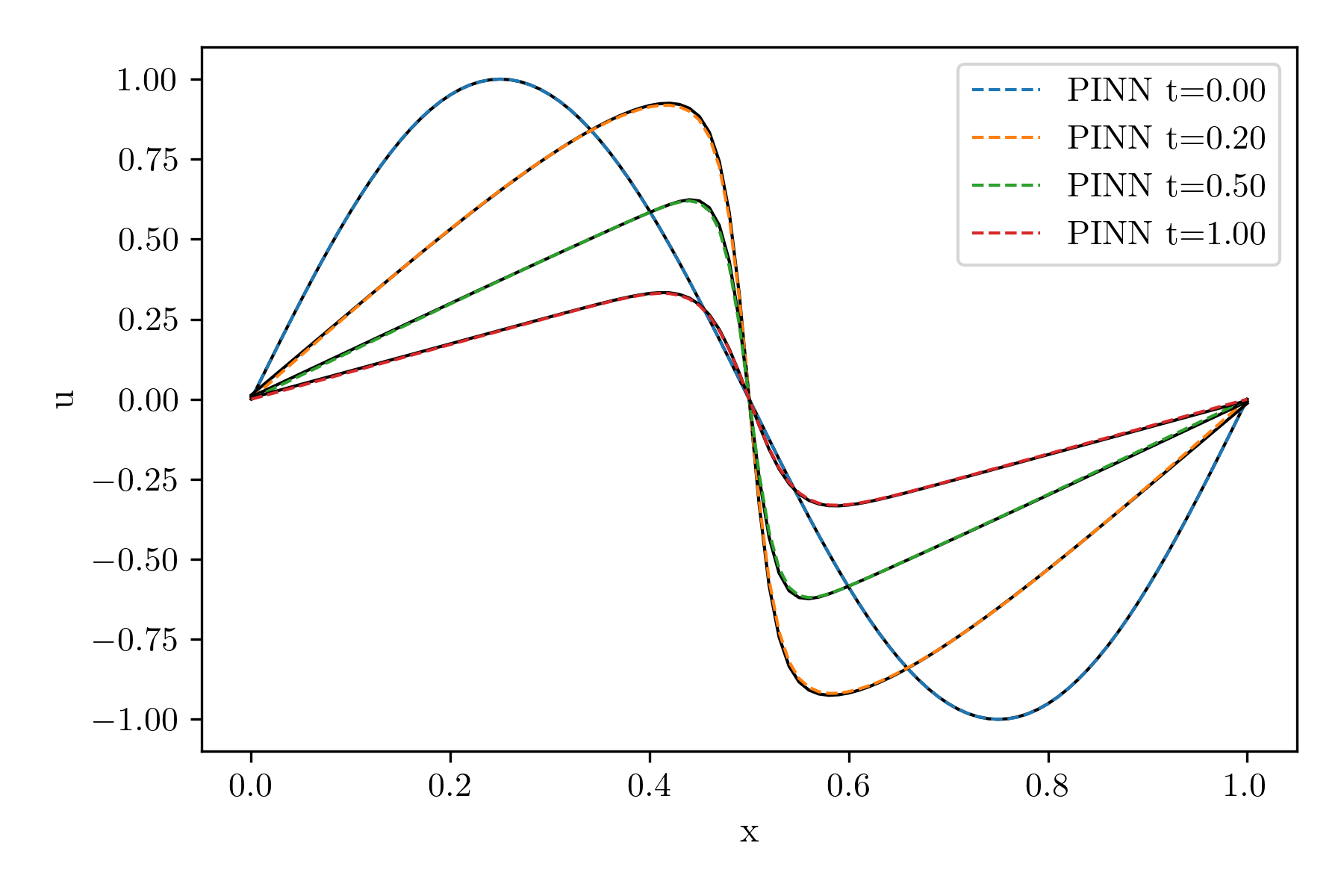} }}
    \caption{Comparison of the solution of Burgers' equation for CFD and DDS-PINN}
    \label{fig:burgercomparison}
\end{figure}

Figure~\ref{fig:burgercomparison}a displays the domain-decomposed results for each temporal subdomain using individual subnetworks (each consisting of 4 hidden layers with 32 neurons). The reconstructed global solution is presented in Figure~\ref{fig:burgercomparison}b, where the characteristic steepening of the wave over time is clearly observable.The framework achieves a physics loss convergence of $O(2.35 \times 10^{-4})$ within $5,000$ epochs. Figure~\ref{fig:burgercomparison}c confirms the strong agreement between the PINN solutions and high-fidelity CFD data—obtained using an in-house solver employing a fifth-order Weighted Essentially Non-Oscillatory (WENO5) scheme—at various time intervals. The accurate depiction of the initial sine wave transitioning into a sharp gradient further verifies the efficacy of the domain-decomposed approach in resolving stiff, transient dynamics.

\subsection{Laminar flat plate boundary layer}
\begin{figure}
\centering
% -------- Left Column --------
\begin{minipage}{0.5\textwidth}
    \centering
    \includegraphics[width=\textwidth]{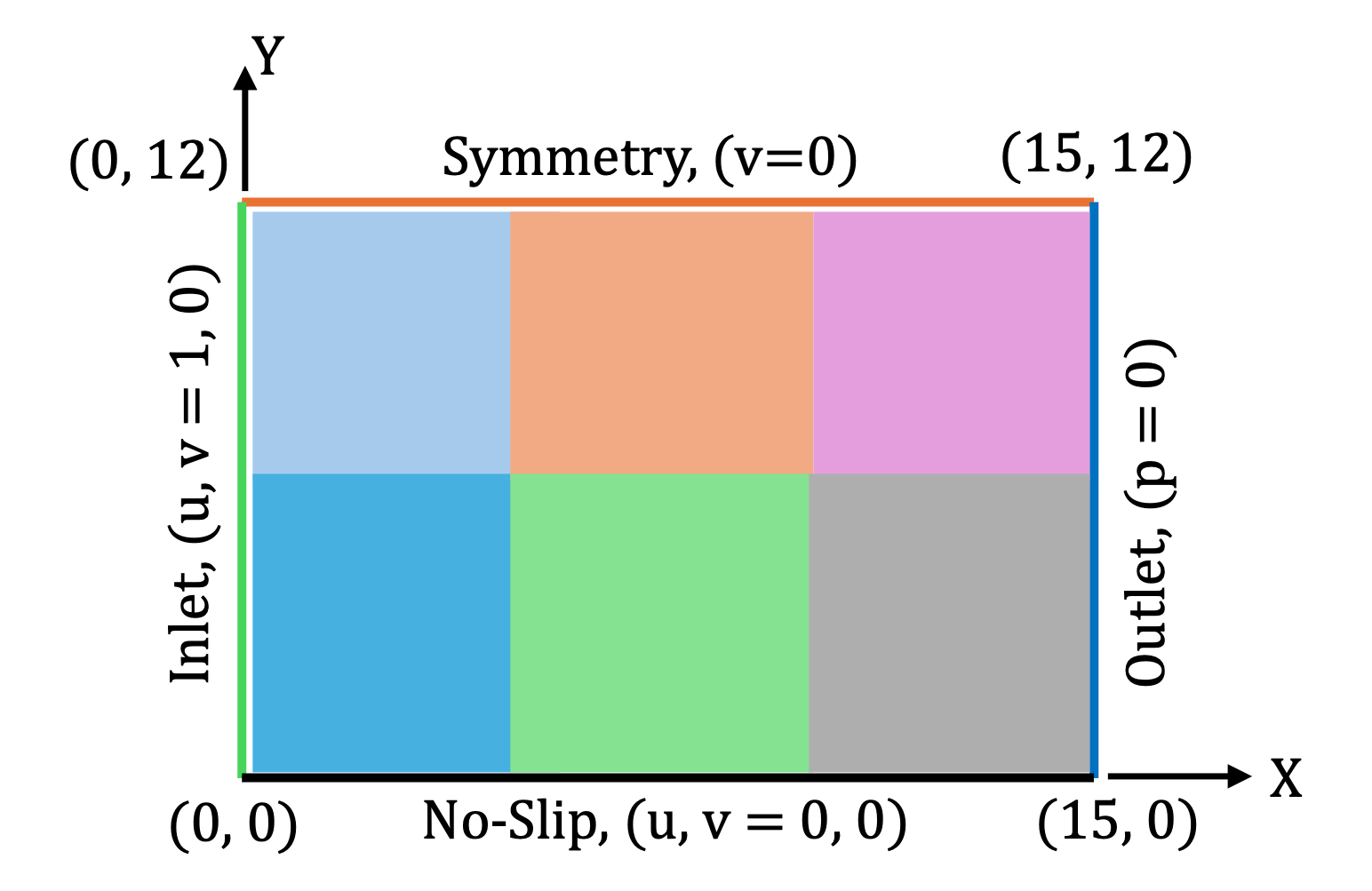}
    \subcaption{Domain-decomposition and boundary condition}
    \vspace{0.5em}
\includegraphics[width=\textwidth]{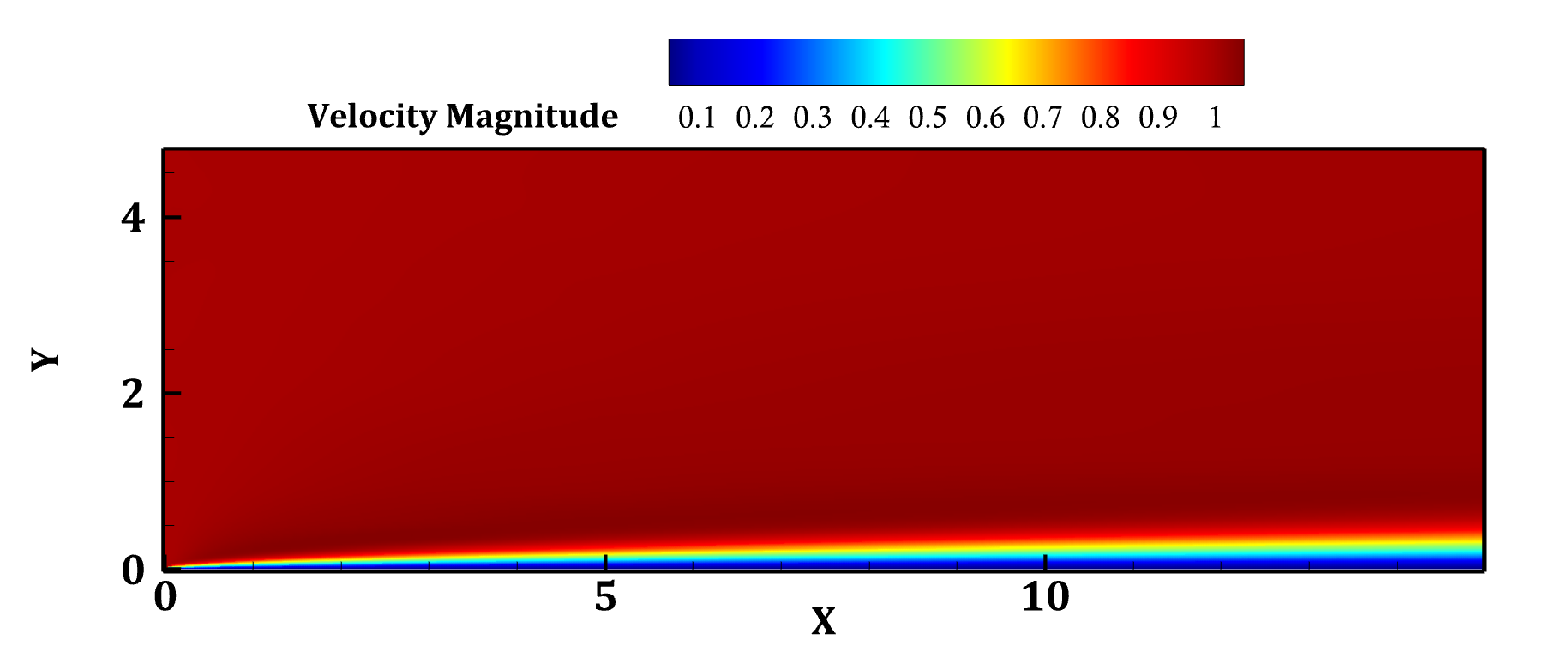}
    \subcaption{Velocity-magnitude contour using PINN}
\end{minipage}
\hfill
% -------- Right Column --------
\begin{minipage}{0.48\textwidth}
    \centering
    \includegraphics[width=\textwidth]{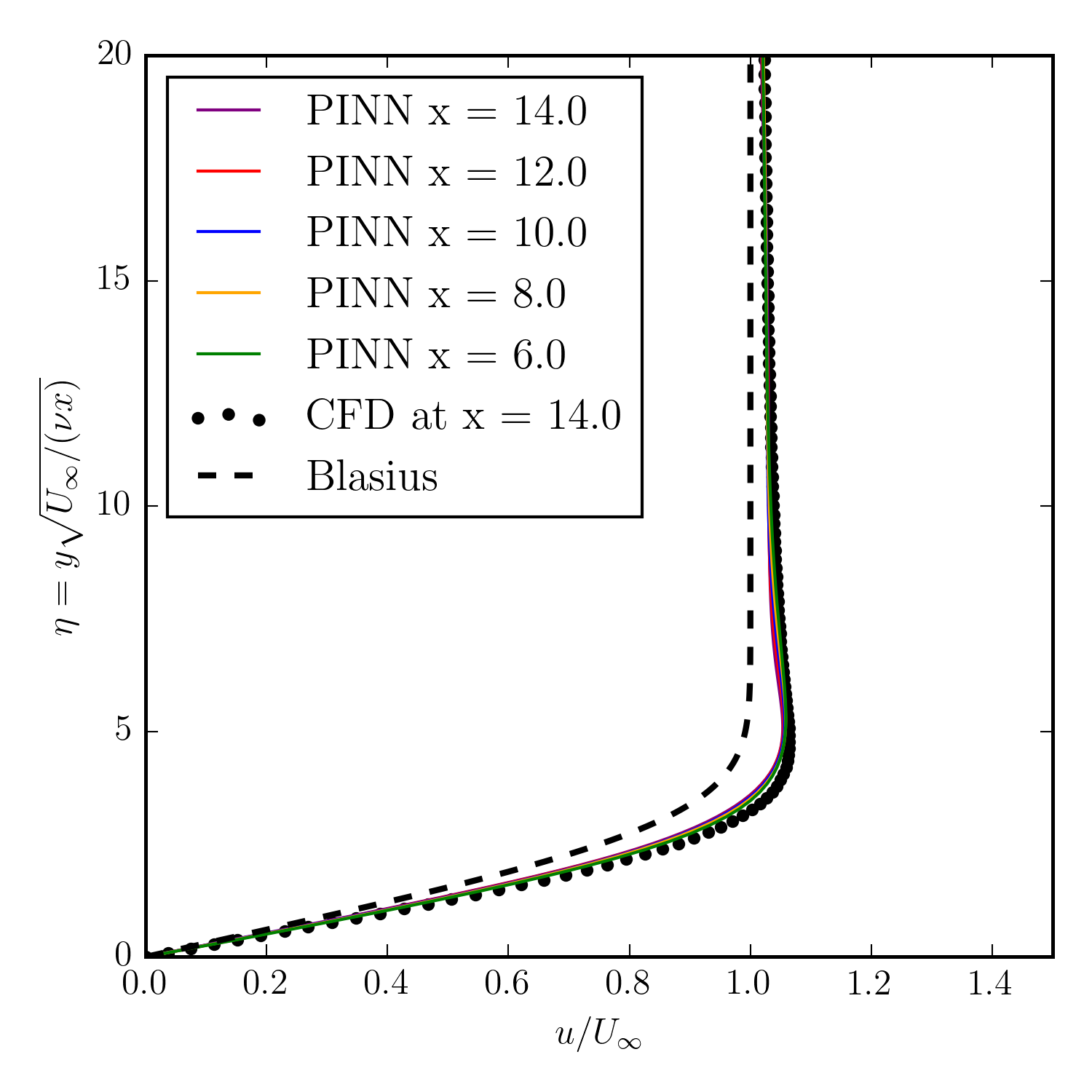}
    \subcaption{Blasius self-similar solution comparison}
\end{minipage}
\caption{Flat plate boundary layer at $Re = 500$}
\label{fig:blasius}
\end{figure}

We consider another benchmark problem in fluid dynamics: the development of a boundary layer over a flat plate. Because the boundary layer is extremely thin near the wall, this presents a significant multiscale challenge for global optimizers like traditional PINNs. The computational domain is a rectangular region defined by $x \in [0,15]$ and $y \in [0,12]$, as shown in Figure~\ref{fig:blasius}a. The governing equations are the two-dimensional steady Navier-Stokes equations, with the Reynolds number set to $Re = 500$ based on unit length.The boundary conditions are prescribed as follows: the left boundary serves as a velocity inlet with $(u,v) = (1,0)$; a no-slip condition $(u,v) = (0,0)$ is imposed along the bottom wall; the top boundary is treated with a symmetry condition; and the right boundary is defined as a pressure outlet with $p = 0$.

For training, we employ $2,000$ collocation points on each boundary to enforce the boundary conditions and $120,000$ interior points to evaluate the physics loss, utilizing mini-batches of $12,000$ points. The physics loss is formulated using an RBA-type $L_2$ loss to adaptively weight the residuals. The computational domain is decomposed into a $(3 \times 2)$ grid of subdomains, uniformly distributed in the streamwise ($x$) and wall-normal ($y$) directions, as indicated by the dashed lines in Figure~\ref{fig:blasius}a. In the DDS-PINN framework, the shift vectors $\mathbf{S}_i$ are set to the geometric centers of their respective subdomains to ensure optimal numerical conditioning.

Each subdomain is associated with an individual neural network comprising five hidden layers with 96 neurons per layer, employing the residual-based attention mechanism described earlier. The model was trained for $25,000$ epochs, during which the physics loss converges to $6.7 \times 10^{-6}$. Figure~\ref{fig:blasius}b presents the velocity magnitude contours, highlighting the thin boundary layer predicted by the PINN; the domain is truncated vertically to emphasize near-wall dynamics.

Figure~\ref{fig:blasius}c illustrates the PINN predictions for the boundary layer profiles at various streamwise locations using similarity scaling. The streamwise velocity ($u$) and wall-normal distance ($y$) are normalized as follows: 
\begin{equation}
    f'(\eta) = \frac{u}{U_\infty}, \quad \eta = y \sqrt{\frac{U_\infty}{\nu x}}
\end{equation}

These scales are chosen to assess the self-similar behavior inherent in the laminar boundary layer. As demonstrated in Figure~\ref{fig:blasius}c, the predicted velocity profiles collapse onto a single curve, exhibiting the expected self-similarity. These results also show excellent agreement with CFD solutions obtained using the commercial software ANSYS Fluent with a pressure-based solver. We also compare the results against the theoretical Blasius solution. Although the overall match is good, the slight difference from computational results is attributed to the relatively low Reynolds number ($Re=500$), as the classical Blasius derivation assumes the high-$Re$ limit where the boundary layer equations are most accurate.

\section{Backward-facing step}
\begin{figure}
\centering
    \subfloat[\centering Computational domain.]
        {{\includegraphics[width=0.86\textwidth]{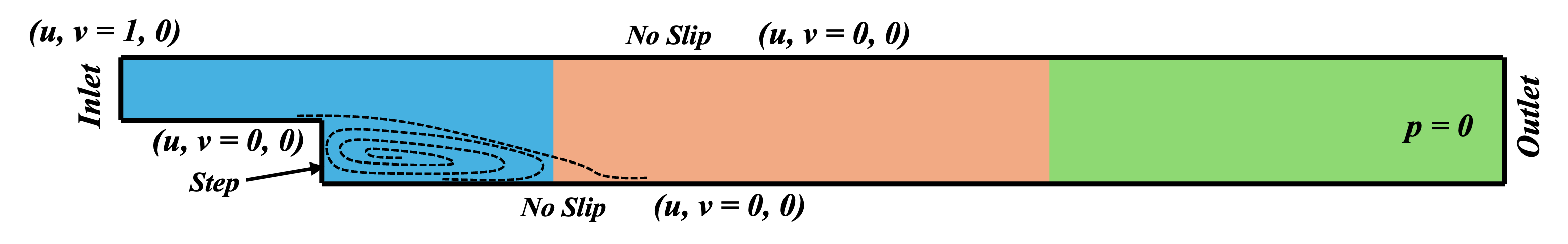} }}\\
           \subfloat[\centering Residual and boundary points distribution. Red dots show data points used for turbulent BFS]
        {{\includegraphics[width=0.8\textwidth]{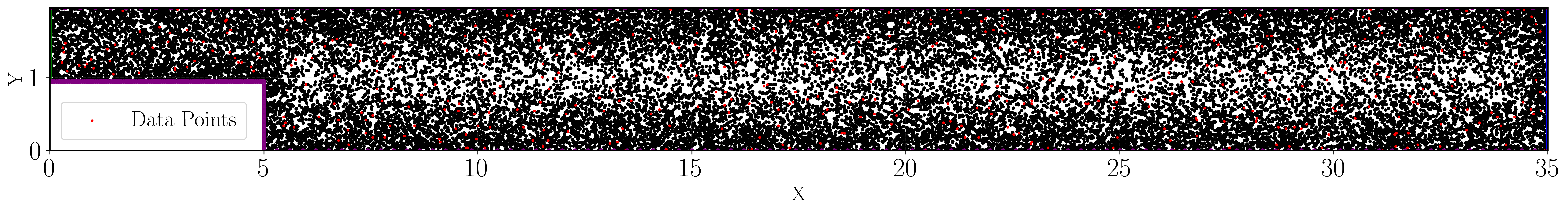} }}
    \caption{Domain and collocation points distribution for backward-facing step (BFS) flow}
    \label{fig:domain}
\end{figure}

The DDS-PINN framework is subsequently applied to the computationally demanding backward-facing step (BFS) problem. Figure~\ref{fig:domain}a illustrates the physical configuration, where the geometric step triggers flow separation and the formation of a localized recirculation region. The characteristic dimensions of this separation zone are inherently dependent on the Reynolds number ($Re$). As shown in the figure, the computational domain spans $(x, y) \in [0, 35] \times [0, 1.9423]$, with the step located at $x = 5$ and $y \in [0, 0.9423]$. Boundary conditions, mirroring those used in classical CFD, are also detailed in the figure. A velocity inlet is defined at the inflow, while a pressure outlet is employed at the outflow; all remaining boundaries are treated as no-slip walls.

An expansive downstream domain (35 units) is utilized to ensure that the outlet boundary conditions do not introduce unphysical reflections into the regions of interest. This domain size introduces significant computational complexity. In the absence of dense supervision data, conventional neural architectures often struggle to achieve the convergence necessary for the seamless propagation of physical information from the velocity inlet to the distal pressure outlet.

As shown in Figure~\ref{fig:domain}a, the domain is decomposed into three adjacent subdomains along the streamwise direction: $\Omega_1 \in [0, 12]$, $\Omega_2 \in [12, 24]$, and $\Omega_3 \in [24, 35]$. These are centered using the shift vectors $\mathbf{S}_1 = (6, 0)$, $\mathbf{S}_2 = (18, 0)$, and $\mathbf{S}_3 = (29.5, 0)$, respectively. Each subdomain employs a dedicated subnetwork that takes shifted coordinates as input and predicts the local flow variables. The subnetworks share a unified architecture consisting of five hidden layers with 96 neurons per layer.

Figure~\ref{fig:domain}b shows the residual and boundary point distributions. We utilize $80,000$ collocation points to enforce the residual loss and $5,000$ points along each boundary to prescribe the boundary conditions. To accurately capture near-wall phenomena, such as boundary layer development and recirculation bubbles, the residual collocation points are distributed with a higher density in the near-wall regions. Training is performed using mini-batches of $20,000$ points for the physics loss, formulated as an RBA-type $L_2$ objective. 

In the following sections, we present PINN predictions for both laminar and turbulent flow conditions. The Reynolds numbers, based on the channel height upstream of the step  $h$, are $Re_h = 100$ and $Re_h = 10,000$, respectively. We solve the steady two-dimensional Navier-Stokes equations for the laminar regime, while for the turbulent regime, the Reynolds-Averaged Navier-Stokes (RANS) equations are employed as the physical constraints.

\subsection{Laminar Flow}\label{app:laminarsection}

\begin{figure}
\centering 
\begin{subfigure}{0.59\textwidth}
\centering
\includegraphics[width=1\textwidth]{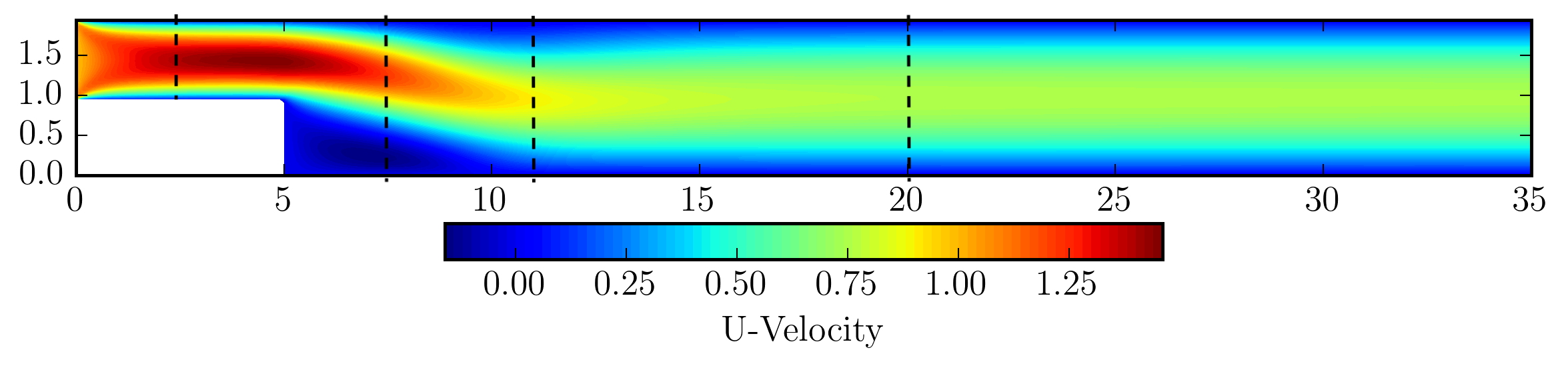}\\[2mm]%
\includegraphics[width=1\textwidth]{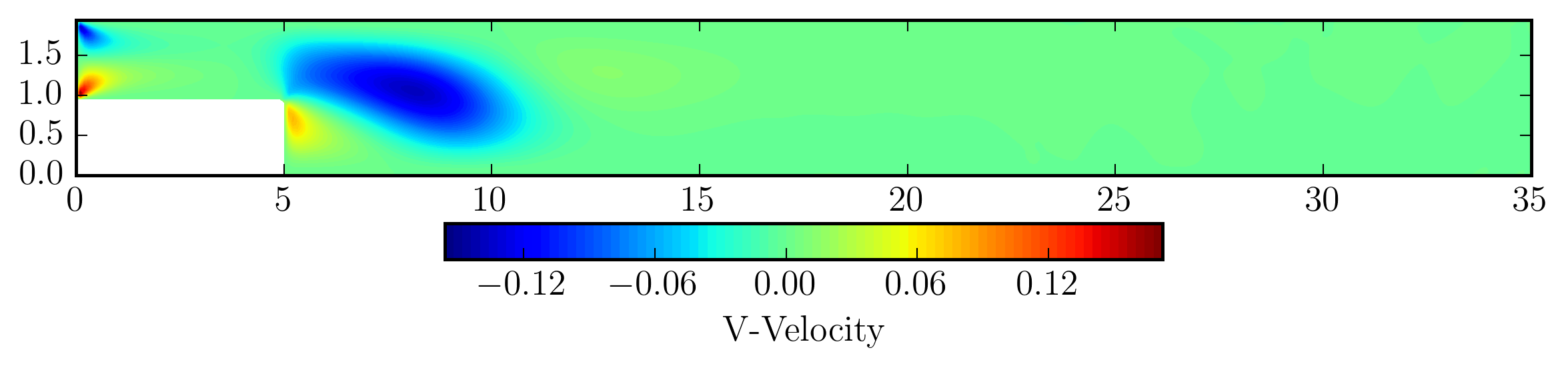} \vfill
\caption{Velocity contours from PINN predictions}\label{fig:left}
\end{subfigure}
\begin{subfigure}{0.4\textwidth}
\centering
\includegraphics[width=1\textwidth, trim=0 0 0 0]{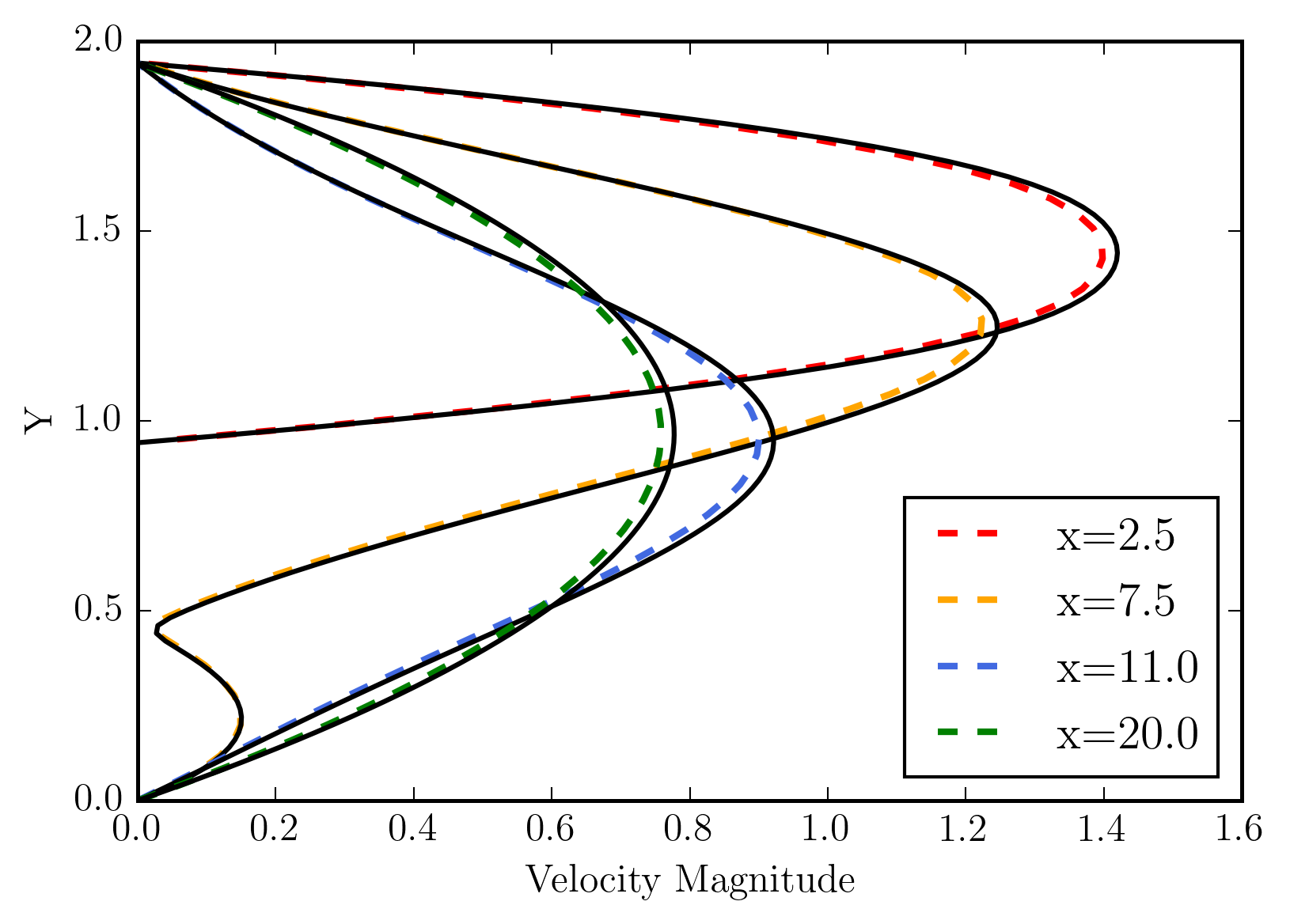}%
\caption{Comparison of PINN (dashed line) and CFD (solid line)}\label{fig:right}
\end{subfigure}
\caption{Predictions of laminar BFS at $Re_h=100$}
\label{fig:wo_data}
\end{figure}

The laminar flow predictions are obtained in a completely data-free manner. In this case, the physics loss converges to $1.5 \times 10^{-5}$, indicating a satisfactory enforcement of the governing equations. Figure~\ref{fig:wo_data}(a) presents the predicted contours of the streamwise ($u$) and wall-normal ($v$) velocity components obtained using the PINN framework, where the recirculation region near the step is characterized by a significant reduction in streamwise velocity.

To verify the accuracy of these results, CFD simulations are performed using ANSYS Fluent, and quantitative comparisons against the PINN framework are conducted at four streamwise stations: $x = 2.5, 7.5, 11.0,$ and $20.0$, as indicated by the dashed vertical lines in Figure~\ref{fig:wo_data}(a). The extracted velocity magnitude profiles are shown in Figure~\ref{fig:wo_data}(b). At all locations, the PINN predictions demonstrate excellent agreement with the CFD data, with relative errors remaining below $5\%$.At the $x = 2.5$ station, the results successfully capture the development of the thin boundary layer upstream of the sudden expansion. At $x = 7.5$ and $x = 11.0$, the velocity magnitude profiles closely match the CFD results, demonstrating that the complex recirculation and reattachment dynamics are accurately resolved. Further downstream at $x = 20.0$, the agreement remains robust, highlighting the ability of the current PINN framework to capture long-range spatial dependencies that originate at the inlet and persist throughout the computational domain.

\subsection{Turbulent Flow}\label{turbsection}

The turbulent BFS problem is solved using the Reynolds-Averaged Navier-Stokes (RANS) equations with the $k\mbox{-}\epsilon$ turbulence model as PDE constraints. 
These equations are detailed in Appendix~\ref{app:turbulent_model}. 
In total, five steady-state equations representing mean continuity, mean momentum, turbulent kinetic energy ($k$), and the dissipation rate ($\epsilon$) are utilized as constraints. 
Consequently, the network predicts five primary flow variables at the output: $(\bar{u}, \bar{v}, \bar{p}, k, \epsilon)$. To ensure physical consistency, $k$ and $\epsilon$ are constrained to positive values using an exponential activation function.

Due to the inherent closure problem and the absence of explicit information on Reynolds stresses within the PDE residuals, sparse supervision data is employed to guide the flow toward convergence. This supervision data is derived from CFD simulations performed on a high-fidelity grid of $170{,}000$ points. From this set, only $500$ points are randomly sampled, and values of mean velocity, pressure and turbulence kinetic energy are incorporated into the data loss term, representing less than $0.3\%$ of the data points in the full domain. The spatial distribution of these sparse supervision points is illustrated in Figure~\ref{fig:domain}(b).

\begin{figure}
\centering
         \subfloat[\centering RBA-PINN]
        {{\includegraphics[width=0.48\textwidth]{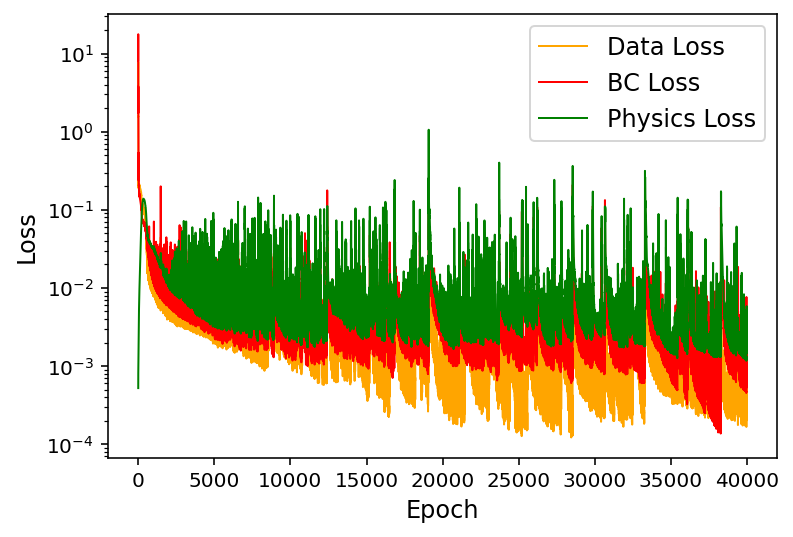} }}
        \subfloat[\centering DDS-RBA PINN]
        {{\includegraphics[width=0.48\textwidth]{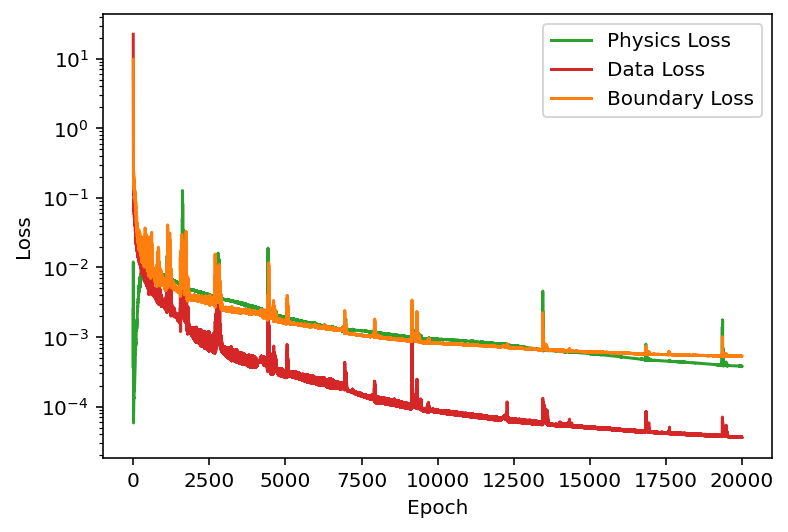} }}
    \caption{Convergence in  turbulent BFS flow predictions at $Re_h = 10{,}000$}
    \label{fig:loss}
\end{figure}

The predictions are obtained using both a single-network architecture (referred to as RBA-PINN) and the DDS-PINN framework. The RBA-PINN architecture consists of 5 hidden layers with 192 neurons each, while all other training hyperparameters remain identical to those employed in DDS-PINN. Figure~\ref{fig:loss} illustrates the loss landscape for both frameworks. In the DDS-PINN framework, the physics loss converges to $\mathcal{O}(10^{-4})$ within $10,000$ epochs. In contrast, the RBA-PINN converges significantly more slowly, reaching only $\mathcal{O}(10^{-3})$ after $40,000$ epochs. These results demonstrate that DDS-PINN achieves approximately four times the convergence speed alongside an order-of-magnitude improvement in accuracy. Despite the reduced number of training iterations, DDS-PINN consistently delivers superior predictive performance and computational efficiency compared to the single-network RBA-PINN.  Loss curves further highlight the stability advantages of the DDS-PINN approach. While the RBA-PINN exhibits significant oscillations in the loss history, the DDS-PINN formulation yields a much smoother and more stable convergence trajectory. 

\begin{figure}
\centering
     \subfloat[\centering RANS CFD (Baseline)]
        {{\includegraphics[width=0.48\textwidth]{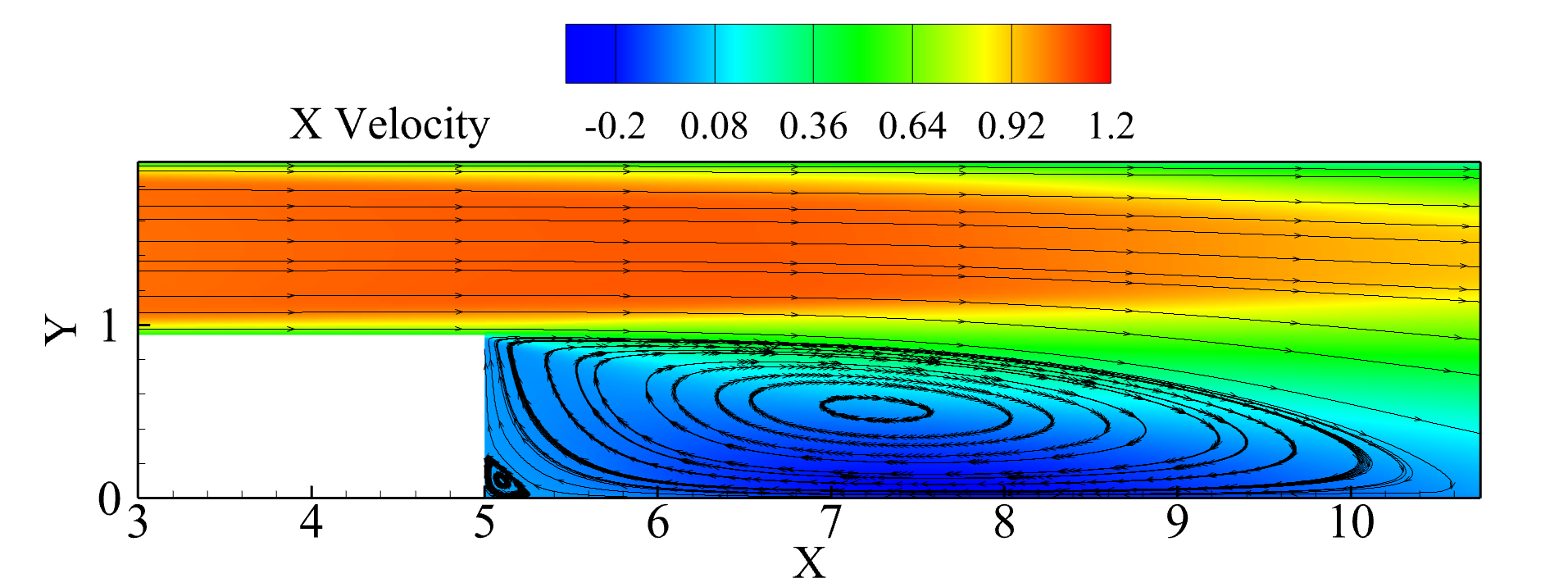}} 
        {\includegraphics[width=0.48\textwidth]{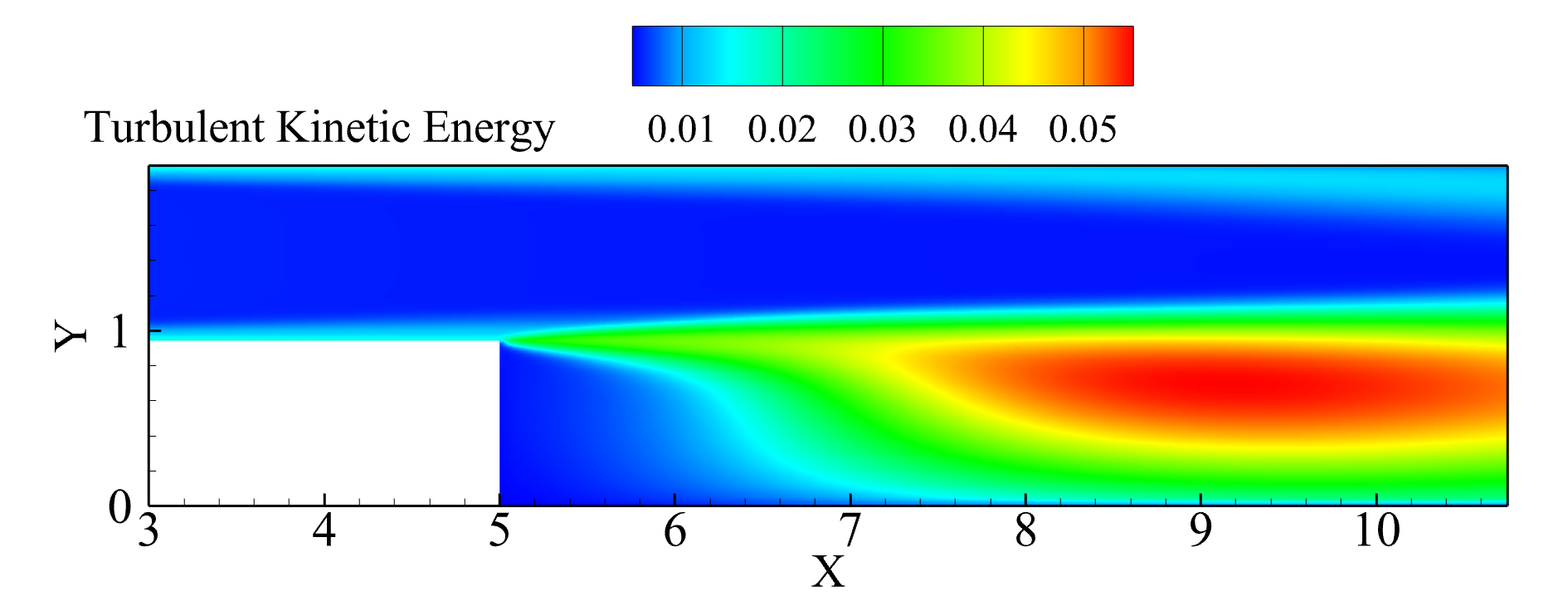} }
        }
        \\
    \subfloat[\centering RBA-PINN (Single wetwork)]
        {{\includegraphics[width=0.48\textwidth]{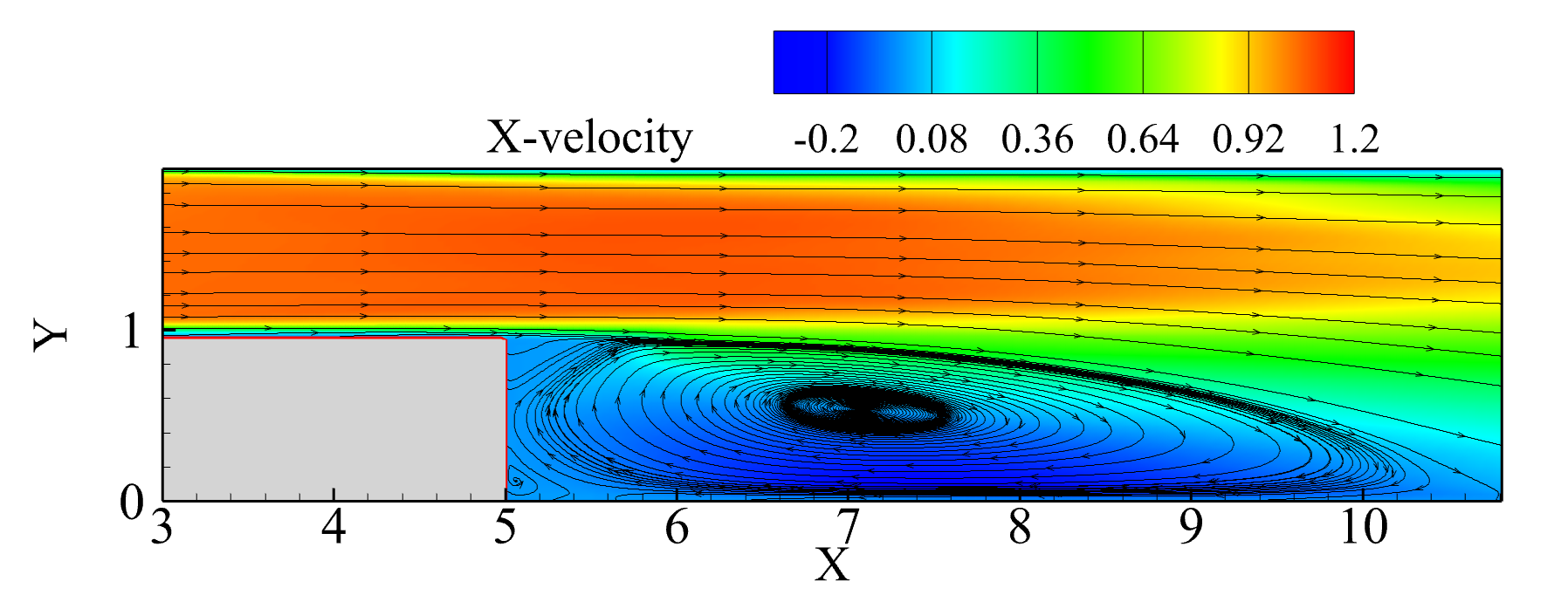} }
        {\includegraphics[width=0.48\textwidth]{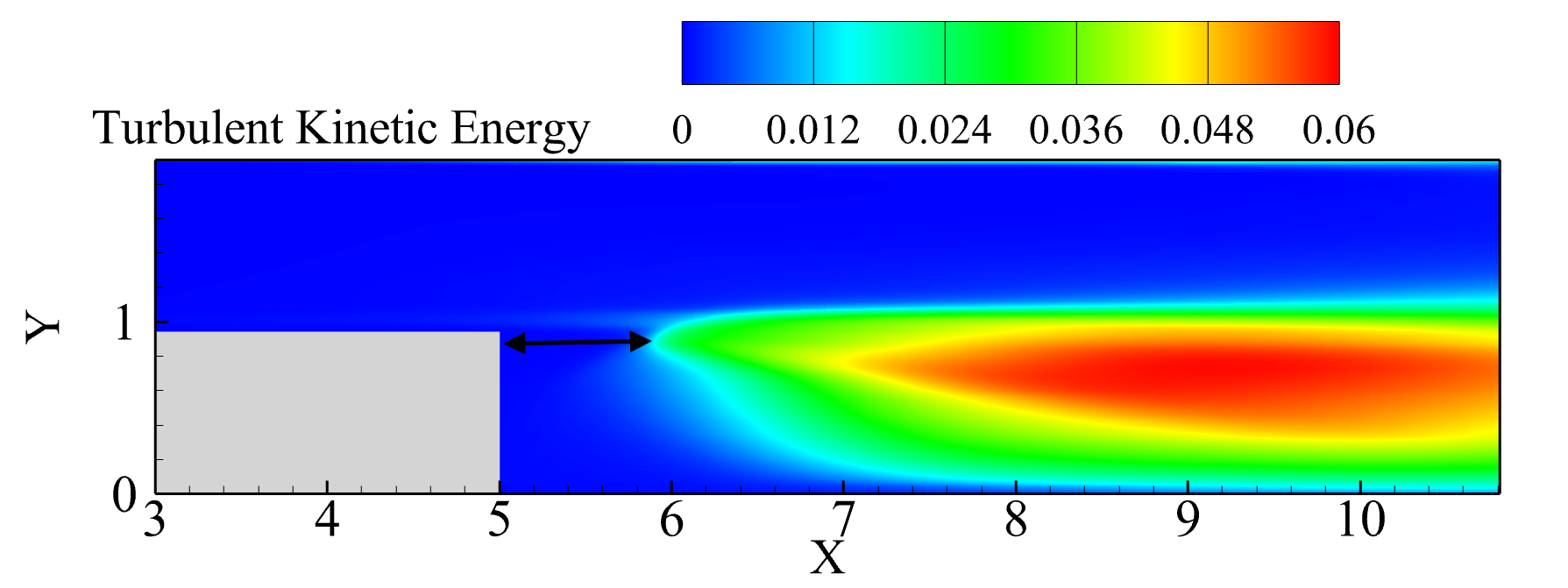} }
        }
        \\
         \subfloat[\centering DDS-PINN]
        {{\includegraphics[width=0.48\textwidth]{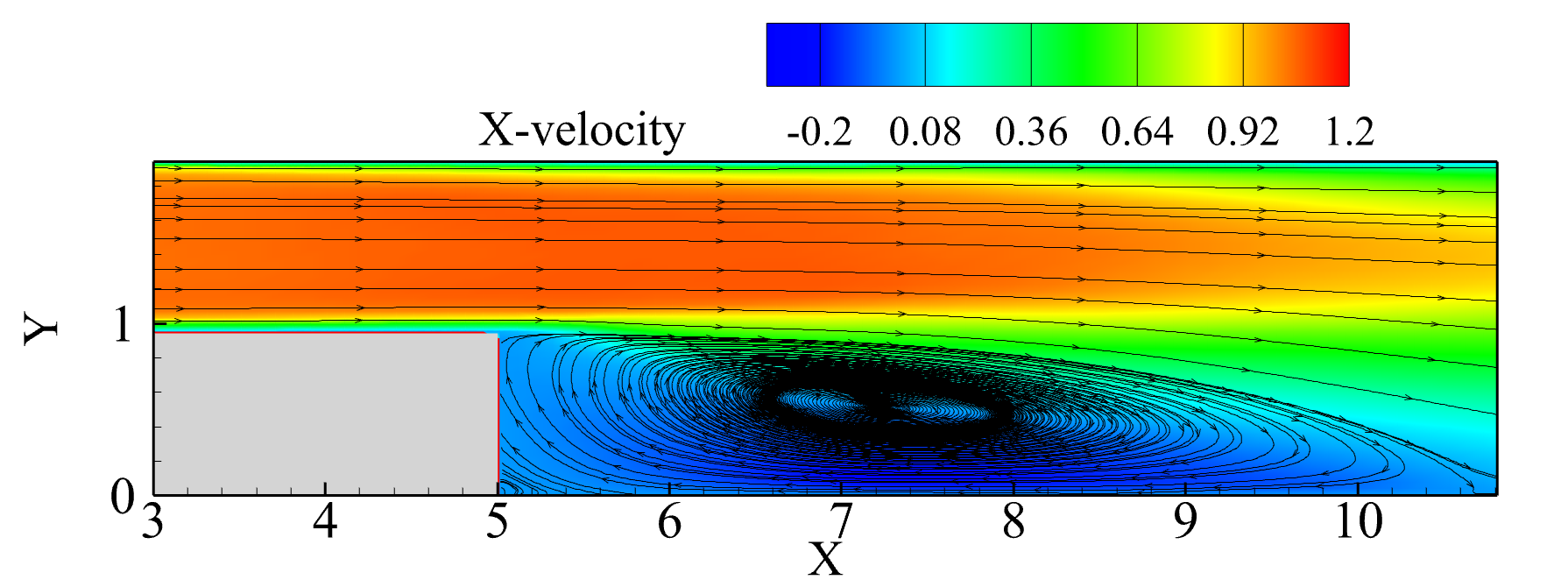} }
        {\includegraphics[width=0.48\textwidth]{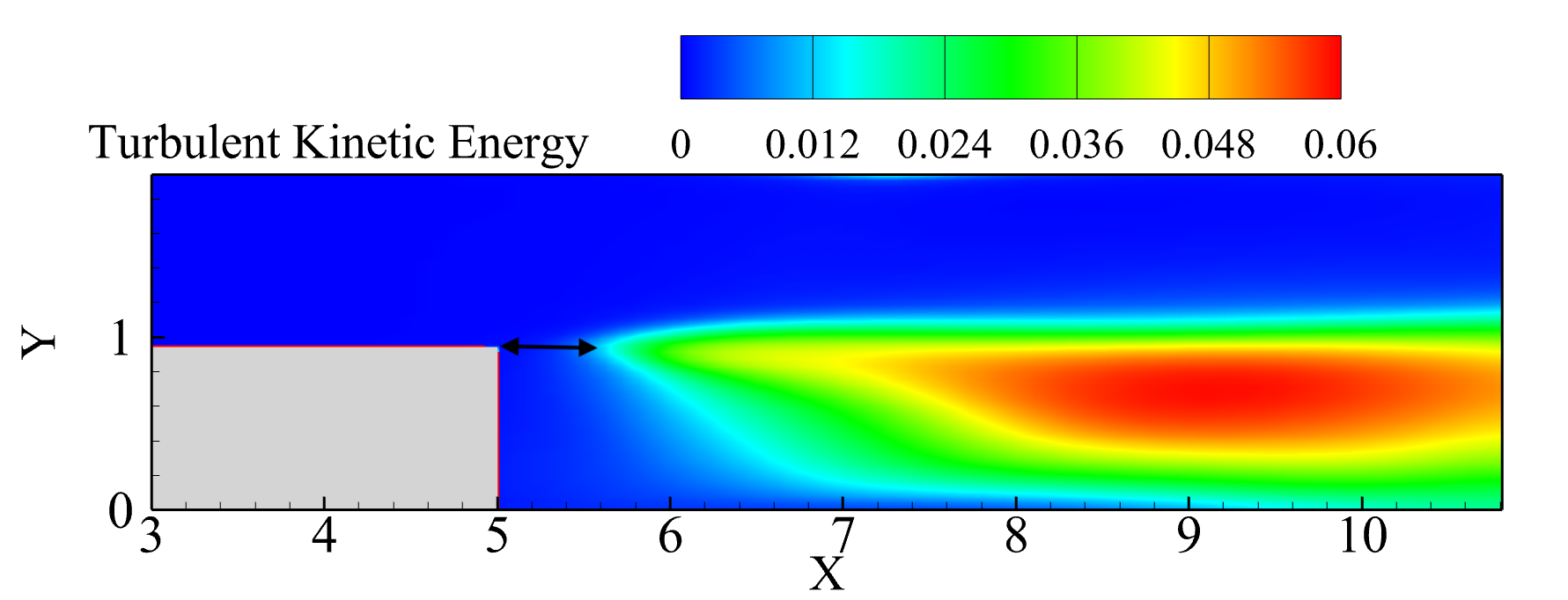} }}
    \caption{Turbulent BFS flow predictions. Left: Streamlines with $\bar{u}$-contours, Right: Turbulence kinetic energy ($k$).}
    \label{fig:Streamlineturb}
\end{figure}

Figure~\ref{fig:Streamlineturb} compares the flow features predicted by RBA-PINN and DDS-PINN against the corresponding high-fidelity CFD benchmarks. The left panel displays streamlines near the separation region overlaid on streamwise velocity contours, while the right panel illustrates the turbulent kinetic energy ($k$) distribution. Although the global contour plots from both PINN architectures appear superficially similar, a detailed examination reveals significant disparities in resolving localized flow structures. Notably, the separation bubble predicted by DDS-PINN aligns more closely with the CFD results than the RBA-PINN prediction. At this Reynolds number, the CFD simulation identifies a secondary recirculation bubble near the bottom corner of the step. This feature is successfully captured by both RBA-PINN and DDS-PINN, marking a distinct improvement over previous studies~\cite{pioch2023turbulence} where turbulence models often failed to resolve this corner vortex. However, the RBA-PINN exhibit significant boundary leakage, characterized by streamlines unphysically penetrating the solid walls. This is a well-documented challenge in PINN literature~\cite{pioch2023turbulence,sun2020surrogate}, arising when the no-slip boundary condition is treated as a soft constraint within a multi-objective loss function. In contrast, the DDS-PINN framework effectively mitigates this issue; the localized training and subdomain-specific coordinate shifting significantly reduce the vertical flux at the walls, leading to a more physically consistent representation of the near-wall physics. Regarding the $k$-contours, both PINN frameworks approximate the global distribution; however, a reduction $k$ is observed downstream of the geometric discontinuities, as indicated by the black arrows. Nevertheless, the DDS-PINN predictions exhibit much closer agreement with the CFD data even in this region.

\begin{figure}
\centering 
\begin{subfigure}{0.59\textwidth}
\centering
\includegraphics[width=1\textwidth]{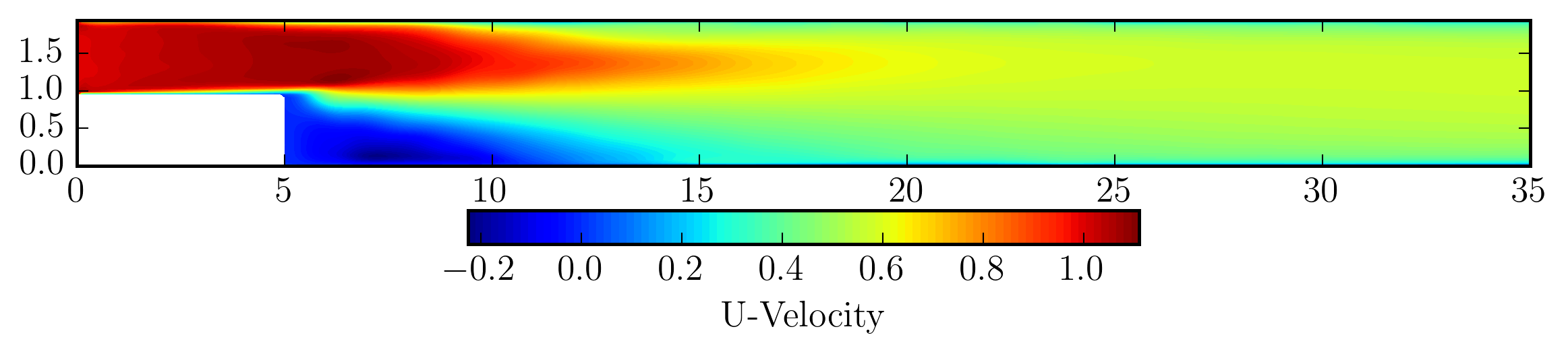}\\[2mm]%
\includegraphics[width=1\textwidth]{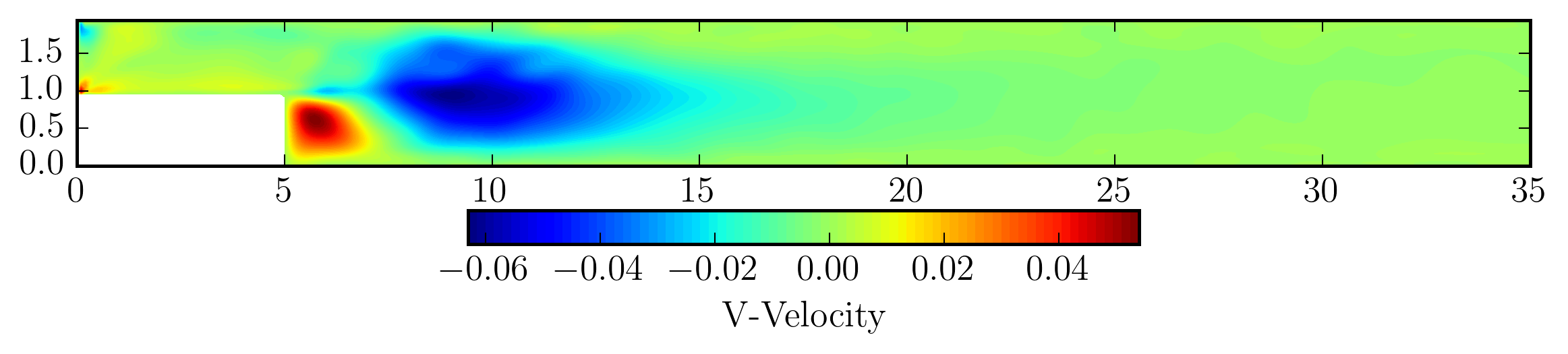} \vfill
\caption{Velocity contours from PINN predictions}\label{fig:left}
\end{subfigure}
\begin{subfigure}{0.4\textwidth}
\centering
\includegraphics[width=1\textwidth, trim=0 0 0 0]{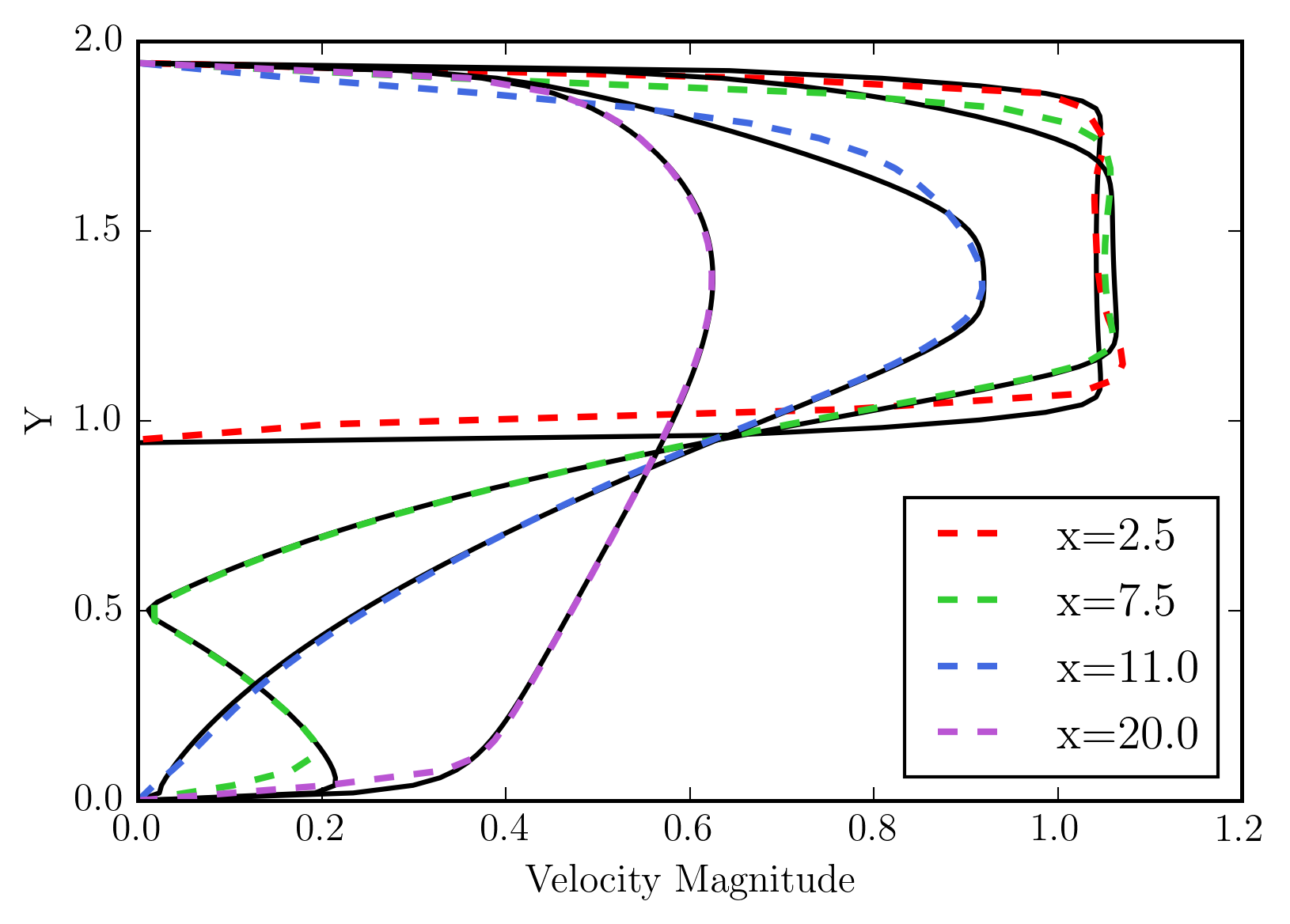}%
\caption{Comparison of DDS-PINN (dashed line) and CFD (solid line)}\label{fig:right}
\end{subfigure}
\caption{Predictions of turbulent BFS at $Re=10{,}000$}
\label{fig:turbbfs}
\end{figure}

Finally, we provide a quantitative comparison between the DDS-PINN predictions and the reference CFD results. Figure~\ref{fig:turbbfs}a displays the predicted contours of the mean streamwise ($\bar{u}$) and wall-normal ($\bar{v}$) velocity components across the entire computational domain. 
The wall-normal velocity becomes negative immediately downstream of the step as the flow expands into the larger cross-section. Conversely, the localized positive values observed in the immediate vicinity of the step are attributed to the secondary corner vortex.
Figure~\ref{fig:turbbfs}b compares the predicted velocity profiles with CFD data at five selected streamwise stations. The PINN predictions demonstrate close agreement with the CFD results, with only minor discrepancies observed in the near-wall regions. At the $x = 2.5$ station, the development of the boundary layer atop the step is accurately captured. Within the recirculation region at $x = 7.5$, the characteristic inflectional velocity profile is clearly observed, though slight deviations from the CFD data persist very close to the wall. In the far-downstream region ($x = 20$), the velocity profiles nearly overlap with the CFD results. Table~\ref{tab:l2error} also shows that the mean square error (MSE) defined as $\frac{1}{n}\sum_{i=1}^{n}\left(V_{mag}^{\text{Predicted}} - V_{mag}^{\text{CFD}}\right)^2$, is highest near the inflow due to the high-gradient boundary layer and decreases downstream, indicating the model’s ability to capture long-range dependencies. Despite these minor differences, the performance of DDS-PINN is highly encouraging, given the complexity of the flow, its inherent multiscale characteristics, and the long-range dependencies between the inlet and outlet. Achieving such accuracy with minimal supervision data underscores the potential of this framework for solving practical, high-Reynolds-number fluid dynamics problems.

\begin{table}
    \centering
\caption{Mean square error (MSE) of mean velocity magnitude at streamwise locations for BFS turbulent flow}
\label{tab:l2error}
    \begin{tabular}{|c|c|c|c|c|}\hline
         Streamwise position&  $x = 2.5$&  $x= 7.5$&  $x= 11.0$& $x= 20.0$\\\hline
         $MSE$&  0.02&  0.002&  0.0016& 0.0002\\ \hline
    \end{tabular}
\end{table}

\section{Conclusions}
In this work, a Physics-Informed Neural Network (PINN) framework is proposed to address the nonlinearity, multiscale features, and long-range dependency issues inherent in complex fluid-dynamics problems. The large computational domain is decomposed into overlapping subdomains, where each subnetwork is supplied with inputs shifted to the geometric center of its corresponding subdomain. Furthermore, a Residual-Based Attention (RBA) strategy is integrated to enhance the resolution of high-gradient regions. This strategy, referred to as Domain-Decomposed and Shifted (DDS)-PINN, demonstrates enhanced convergence, accuracy, and computational efficiency for multiscale ODEs compared to vanilla PINNs and other relevant frameworks in the literature. DDS-PINN is further employed to solve the unsteady Burgers’ equation, laminar flow over a flat plate, and laminar backward-facing step (BFS) flow in a completely data-free manner, yielding results comparable to traditional CFD simulations. The proposed approach is also applied to simulate turbulent flow over a backward-facing step at $Re_h = 10,000$. The results indicate that DDS-PINN, even when trained with a limited number of sparse supervision points (less than $0.3\%$ of the domain), produces predictions that are consistently in agreement with high-fidelity CFD solutions. In particular, DDS-PINN successfully captures fine-scale flow features—such as boundary layers and secondary separation bubbles—that are not adequately resolved by single-network architectures. Owing to its scalability, improved accuracy, and accelerated convergence, DDS-PINN shows strong potential as a robust modeling framework for simulating complex and turbulent fluid-dynamical problems. Further, beyond its utility as a high-fidelity PDE solver, the DDS-PINN framework may function as a potent physics-informed data-enhancement tool for experimental fluid mechanics. By leveraging the underlying governing equations (RANS or Navier-Stokes) as a regularizer, the framework can reconstruct a continuous, high-fidelity flow field from extremely sparse and localized experimental probes.

\section*{Data Availability}
The data that supports the findings of this study are available from the corresponding author upon reasonable request.

\section*{Conflict of Interest}
The authors have no conflicts to disclose.
\section*{Acknowledgement}
The authors acknowledge Ms. Muskan Tongaria for her contributions to the initial RBA-PINN simulations. This work was partially supported by the Prime Minister’s Research Fellowship (PMRF), India (Ref. No. PMRF-2303406) and the Anusandhan National Research Foundation (ANRF) Advanced Research Grant (Grant No. ANRF/ARG/2025/002260/ENS). The computational simulations were performed using the high-performance computing (HPC) facilities at the Kotak School of Sustainability (KSS), IIT Kanpur, and the GPU resources at the Aeroscience Computations and Analysis Laboratory (ACAL).

\small

\bibliography{references}

% %%%%%%%%%%%%%%%%%%%%%%%%%%%%%%%%%%%%%%%%%%%%%%%%%%%%%%%%%%%%

\appendix

\section{2D steady, Incompressible RANS equations with $k\mbox{-}\epsilon$ turbulence model}\label{app:turbulent_model}
Reynolds-Averaged Navier-Stokes (RANS) equations for 2D steady, incompressible flow solve mean flow variables, including pressure ($\overline{p}$) and velocities ($\overline{u}, \overline{v}$). The equations are based on the principles of mass and momentum conservation, and are given below:

\begin{eqnarray}
\text{Continuity:} \quad &&
\frac{\partial \overline{u}}{\partial x} + \frac{\partial \overline{v}}{\partial y} = 0 \\[8pt]
\text{X-momentum:} \quad &&
 \overline{u} \frac{\partial \overline{u}}{\partial x} + \overline{v} \frac{\partial \overline{u}}{\partial y} = 
- \frac{1}{\rho} \frac{\partial \overline{p}}{\partial x}
+ \frac{\partial}{\partial x} \left[ \left( \nu + \nu_t \right) \frac{\partial \overline{u}}{\partial x} \right]
+ \frac{\partial}{\partial y} \left[ \left( \nu + \nu_t \right) \frac{\partial \overline{u}}{\partial y} \right] \\[8pt]
\text{Y-momentum:} \quad &&
 \overline{u} \frac{\partial \overline{v}}{\partial x} + \overline{v} \frac{\partial \overline{v}}{\partial y} = 
- \frac{1}{\rho} \frac{\partial \overline{p}}{\partial y}
+ \frac{\partial}{\partial x} \left[ \left( \nu + \nu_t \right) \frac{\partial \overline{v}}{\partial x} \right]
+ \frac{\partial}{\partial y} \left[ \left( \nu + \nu_t \right) \frac{\partial \overline{v}}{\partial y} \right]
\end{eqnarray}
Where $\nu_t = C_\mu \frac{k^2}{\epsilon}$ is the eddy viscosity, obtained from $k\mbox{-}\epsilon$ turbulence model, where  $k$ is the turbulent kinetic energy, and $\epsilon$ is the dissipation rate. The equations for $k$ and $\epsilon$ are given as:
\begin{eqnarray}
\overline{u} \frac{\partial k}{\partial x} + \overline{v} \frac{\partial k}{\partial y} &=&
\frac{\partial}{\partial x} \left[ \left( \nu + \frac{\nu_t}{\sigma_k} \right) \frac{\partial k}{\partial x} \right]
+ \frac{\partial}{\partial y} \left[ \left( \nu + \frac{\nu_t}{\sigma_k} \right) \frac{\partial k}{\partial y} \right]
+ P_k - \epsilon \\[10pt]
\overline{u} \frac{\partial \epsilon}{\partial x} + \overline{v} \frac{\partial \epsilon}{\partial y} &=&
\frac{\partial}{\partial x} \left[ \left( \nu + \frac{\nu_t}{\sigma_\epsilon} \right) \frac{\partial \epsilon}{\partial x} \right]
+ \frac{\partial}{\partial y} \left[ \left( \nu + \frac{\nu_t}{\sigma_\epsilon} \right) \frac{\partial \epsilon}{\partial y} \right]
+ C_{\epsilon 1} \frac{\epsilon}{k} P_k - C_{\epsilon 2} \frac{\epsilon^2}{k}
\end{eqnarray}

Where $P_k$ is the production of turbulent kinetic energy, given as:
\begin{align}
P_k &= 2\mu_t\left[\left(\frac{\partial \overline{u}}{\partial x}\right)^2 + \left(\frac{\partial \overline{v}}{\partial y}\right)^2 + \frac{1}{2}\left(\frac{\partial \overline{u}}{\partial y} + \frac{\partial \overline{v}}{\partial x}\right)^2\right] 
\end{align}

The model parameters are taken as below~\cite{launder1983numerical}:
\begin{align}
C_\mu = 0.09, \quad \sigma_k &= 1.0, \quad \sigma_\epsilon = 1.3, \quad C_{\epsilon 1} = 1.44, \quad C_{\epsilon 2} = 1.92
\end{align}

\end{document}